\documentclass[12pt,a4paper]{article} 
\usepackage{amsmath}
\usepackage{graphicx}
\usepackage{bm}
\usepackage{hyperref,xcolor}
\definecolor{colorlink}{rgb}{0.2,0.2,0.2}
\hypersetup{colorlinks,linkcolor=colorlink,citecolor=blue}
\usepackage{authblk}

\renewcommand{\slash}[1]{%
  \mathrel{\setbox0=\hbox{$/$}\copy0\kern-1.1\wd0\hbox{$#1$}}}
\newcommand{\Slash}[1]{%
\mathrel{\setbox0=\hbox{$#1$}\copy0\kern-0.75\wd0\hbox{$/$}}}
\newcommand\TT{\rule{0pt}{2.5ex}}        
\newcommand\BB{\rule[-1.0ex]{0pt}{0pt}}  

\newcommand{\be}{\begin{equation}}
\newcommand{\en}{\end{equation}}
\newcommand{\bea}{\begin{eqnarray}}
\newcommand{\ena}{\end{eqnarray}}
\newcommand{\bp}{\begin{pmatrix}}
\newcommand{\ep}{\end{pmatrix}}
\newcommand{\lbl}[1]{\label{eq:#1}}

\newcommand{\lblsect}[1]{\label{sec:#1}}
\newcommand{\rf}[1]{(\ref{eq:#1})}

\newcommand{\fig}[1]{\ref{fig:#1}}
\newcommand{\sect}[1]{\ref{sec:#1}}
\newcommand{\braque}[1]{{\langle #1 \rangle}}

\newcommand{\bc}{\begin{center}}
\newcommand{\ec}{\end{center}}
\newcommand{\bt}{\begin{tabular}}
\newcommand{\et}{\end{tabular}}
\newcommand{\ba}{\begin{array}}
\newcommand{\ea}{\end{array}}
\newcommand{\undemi}{{1/2}}
\newcommand{\trdemi}{{3/2}}

\newcommand{\gapprox}{%
\mathrel{%
\setbox0=\hbox{$>$}\raise0.6ex\copy0\kern-\wd0\lower0.65ex\hbox{$\sim$}}}
\newcommand{\lapprox}{%
\mathrel{%
\setbox0=\hbox{$<$}\raise0.6ex\copy0\kern-\wd0\lower0.65ex\hbox{$\sim$}}}
\newcommand{\inleft}{%
\mathrel{%
\setbox0=\hbox{$<$}\copy0\kern-0.5\wd0\lower1.1\ht0\hbox{$\scriptstyle{in}$}}}
\newcommand{\inright}{%
\mathrel{%
\setbox0=\hbox{$>$}\copy0\kern-0.5\wd0\lower1.1\ht0\hbox{$\scriptstyle{in}$}}}
\newcommand{\outleft}{%
\mathrel{%
\setbox0=\hbox{$<$}\copy0\kern-0.5\wd0\lower1.1\ht0\hbox{$\scriptstyle{out}$}}}
\newcommand{\outright}{%
\mathrel{%
\setbox0=\hbox{$>$}\copy0\kern-0.5\wd0\lower1.1\ht0\hbox{$\scriptstyle{out}$}}}

\newcommand{\im}{{\rm Im\,}}
\newcommand{\re}{{\rm Re\,}}
\newcommand{\disc}{{\rm disc\,}}
\newcommand{\Kbar}{\bar{K}}
\newcommand{\mpi }{m_\pi}
\newcommand{\mk }{m_K}
\newcommand{\fpi }{F_\pi}

\newcommand{\mpid}{m_\pi^2}

\newcommand{\fpid}{F_\pi^2}

\newcommand{\mkd}{m_K^2}

\newcommand{\Kp}{{K^+}} 

\newcommand{\Kz}{{K^0}} 

\newcommand{\piz}{{\pi^0}} 
\newcommand{\pip}{{\pi^+}}
\newcommand{\pim}{{\pi^-}}

\begin{document}

\title{A dispersive approach to the CP conserving $K\to\pi
  \ell^+\ell^-$ radiative decays }

\author[a]{V\'eronique Bernard}

\author[a]{S\'ebastien Descotes-Genon}

\author[b]{Marc Knecht}

\author[a]{Bachir Moussallam}

\affil[a]{\small Universit\'e Paris-Saclay, CNRS/IN2P3, IJCLab, 91405 Orsay, France}

\affil[b]{\small CNRS/Aix-Marseille Univ./Univ. de
  Toulon,  Centre de Physique Th\'eorique (UMR 7332), CNRS-Luminy Case 907, 13288 Marseille Cedex 9, France}

\maketitle

\begin{abstract}
We reconsider the constraints on the form factors $W_+ (s)$ and $W_S (s)$, 
describing the radiative decay modes $K^+\to\pi^+ \ell^+\ell^-$ and 
$K_S\to\pi^0 \ell^+\ell^-$,  associated with the general  properties of
analyticity and unitarity. Starting from the simple consideration of the
asymptotic behaviours of the two combinations $2 W_+ (s) - W_S (s)$ and 
$W_+ (s) + W_S (s)$, we
derive a minimal pair of dispersive representations which involves only two free
parameters. An important input for these representations consists of the $K\to3\pi$ decay amplitudes, for
which we use a set of solutions of the Khuri-Treiman equations
obtained recently. These
solutions provide an extrapolation from the physical $K\to3\pi$
decay region up to the
resonant $K\pi\to\pi\pi$ scattering regions. 
We show that the experimental energy dependence of $|W_+|^2$ 
can be well reproduced and that the sign of $W_+$ is unambiguously determined.  
We also show that the yet unknown
$\Delta{I}=1/2$ part of the $K_S\to \pi^+ \pi^- \pi^0$ amplitude can 
be determined from the value of
$W_+(0) + W_S(0)$. The possibility of fixing the sign of $W_S(0)$ 
using experimental data on both $|W_+|^2$ and $|W_S|^2$ is discussed.
\end{abstract}

\vspace{0.5cm}
\tableofcontents

\section{Introduction}
The Kaon is the next-to-lightest hadron, yet it is sensitive to
physics at scales considerably larger than its mass, which allowed
for instance the discovery of CP
violation~\cite{Christenson:1964fg}. 
Of particular relevance in this regard are 
the CP-conserving rare radiative decay modes $K^\pm\to\pi^\pm\ell^+\ell^-$
and $K_S\to\pi^0\ell^+\ell^-$. The latter amplitude
is of interest because it makes an indirect CP-violating
contribution to the $K_L \to\pi^0\ell^+\ell^-$ amplitude. This
ultra-rare mode which is dominantly induced by short distance local
operators, is part of the ``golden modes'', together with the modes
$K^\pm\to\pi^\pm\nu\bar{\nu}$, $K_L\to\pi^0\nu\bar{\nu}$,
which are able to probe 
the CKM dynamics of the Standard Model (SM)
with a high theoretical accuracy
(see e.g. ref.~\cite{Anzivino:2023bhp} for a summary of recent achievements
in Kaon physics and future plans). A precise evaluation of this indirect
CP violating contribution  to the decay rate requires information on both the modulus and
the phase of the $K_S$ radiative amplitude relative to that of the 
direct CP violating one.

Concerning the charged amplitude $K^\pm\to\pi^\pm\ell^+\ell^-$, several
experiments have performed measurements of  both the 
branching fraction and the  decay
distribution as a function of the dilepton energy $s$, in the $e^+e^-$~\cite{Appel:1999yq,Batley:2009aa} 
 as well as in the $\mu^+\mu^-$ 
mode~\cite{Batley:2011zz,NA62:2022qes}. Comparing the $e^+e^-$ and $\mu^+\mu^-$ amplitudes was shown to provide  constraints
on a class of operators which violate lepton flavour universality~\cite{Crivellin:2016vjc,DAmbrosio:2022kvb}. The most
recent results by the NA62 collaboration are based on a 
sample  of
$28000$ events in the muon mode. The collaboration expects
to collect over 100K events in each of the $\ell^+\ell^-$ modes at the end of the  present run.

In the case of the neutral mode $K_S\to\pi^0\ell^+\ell^-$, in
contrast, only 7 events have been
observed in the $e^+e^-$ mode~\cite{NA481:2003cfm}  and 
6 events in the $\mu^+\mu^-$ mode~\cite{NA481:2004nbc}. New measurements 
are expected to be performed by the LHCb collaboration which should
be able to improve the statistics by a factor of 40 
approximately~\cite{NA62KLEVER:2022nea,AlvesJunior:2018ldo} in 
the $\mu^+\mu^-$ mode. 

Arguments have been proposed which favour a positive sign for the
interference term between the direct and indirect CP-violating
contributions to the $K_L\to\piz\ell^+\ell^-$
amplitude~\cite{Buchalla:2003sj,Friot:2004yr}. More recently, it has been shown~\cite{DAmbrosio:2024ncc} that a positive interference is actually a property of QCD in the limit where the number of colours $N_c$ becomes infinite.
In principle, lattice QCD is the
most appropriate approach for evaluating this sign from first principles, 
but the present results on the radiative amplitudes are not yet sufficiently
precise~\cite{RBC:2022ddw}.  In this context, it would be useful to determine 
both the magnitude and the (appropriately defined\footnote{We will define this phase here relative to the (non-perturbative) 
chiral coupling $G_8$. In order to correctly compute the interference term in the 
$K_L\to\piz\ell^+\ell^-$ branching fraction knowledge of the relative phase
between $G_8$ and the imaginary part of the perturbative parameter $C_{7V}$ 
is also necessary.}) phase of the 
$K_S\to\piz\ell^+\ell^-$ amplitude from experiment, given a 
theoretical framework for analysing the data. The framework that has been  
usually employed is the so-called ``beyond one-loop model''
(B1L) proposed in ref.~\cite{DAmbrosio:1998gur}. As compared with the calculation
in the chiral expansion at order $p^4$~\cite{Ecker:1987qi,Ananthanarayan:2012hu}
the B1L model includes terms of higher chiral order, which are needed in order
to properly describe the experimental energy dependence. The B1L amplitudes are
written as follows
\be\lbl{B1L}
W_{+,S}(s)=G_F\mkd\big( a_{+,S} + b_{+,S}\frac{s}{\mkd} \big) +W^{\pi\pi}_{+,S}(s)\ 
\en
where the parts linear in $s$ involve four parameters 
to be determined from experiment and the functions $W^{\pi\pi}_{+,S}$ account
for the $\pip\pim$ channel in the unitarity relation, inducing a non linear
energy dependence around the $\pi\pi$ threshold.
These unitarity contributions 
(as we will review below)
are associated with the $K^\pm\to\pip\pim\pi^\pm$ 
and the $K_S\to\pip\pim\piz$
amplitudes, respectively. In the B1L model, these are described 
using quadratic expansions in terms of the Dalitz parameters 
$X$, $Y$ determined from the experimental $K\to3\pi$ data (see
the review~\cite{Devlin:1978ye}). 
Performing fits of the experimental data on $|W_+(s)|^2$ using the B1L
model favours, in general, a negative value
for $a_+$ (see~\cite{DAmbrosio:2018ytt}). Concerning $W_S$, from the
available measurements of the branching fractions it is not possible to 
determine $a_S$ and $b_S$ independently without additional assumptions.
The modulus of $a_S$ has been estimated in ref.~\cite{Buchalla:2003sj} 
assuming a vector-dominance type relation $b_S/a_S=\mkd/m_\rho^2$ (the sign
of this ratio agrees with that obtained in the large $N_c$  
calculation of 
ref.~\cite{DAmbrosio:2024ncc} but the magnitude differs by a factor of two). 

Motivated by the continued experimental progress, and the hope to
reduce the number of free parameters, we develop in this paper a
dispersive description of the amplitudes $W_+$, $W_S$, i.e. based on
the general non-perturbative properties of unitarity and analyticity.
A similar approach has been applied
previously to the radiative di-photon amplitude
$K_L\to\piz2\gamma$~\cite{Kambor:1993tv}.  
This is performed in two steps: 1) We use the
$K\to3\pi$ amplitudes provided in ref.~\cite{Bernard:2024ioq}, which
involve a set of 11 independent solutions of the Khuri-Treiman
(KT) integral equations~\cite{Khuri:1960zz,Sawyer:1960hrc}. These
amplitudes have a domain of validity which extends beyond the physical
decay region and can describe the scattering $K\pi\to\pi\pi$ in the
region of the $\rho(770)$ resonance. 2) We include the $K\pi$
channels, in addition to $\pi^+\pi^-$, in the unitarity relation. This
was not done previously, but it proves important, from a formal point
of view, in order to obtain a representation which achieves a
consistent cancellation of the QCD scale dependence upon adding the
long-distance contribution and the short-distance 
one from the local operator $Q_{7V}$ 
at any value of the energy $s$. Considering
the asymptotic behaviour we will show that a pair of dispersive representations
which involves only two parameters (which can be taken as $a_+$, $a_S$)
should hold. Finally, as an effect of $K^+\pim\to K^0\pi^0$
scattering, our representation couples the two amplitudes $W_+$, $W_S$
such that, in principle, information on both $a_+$ and $a_S$ can be
extracted from data on a single amplitude e.g. $W_+$.

The plan of the paper is as follows. We first introduce in
sec.~\sect{2} the notation and the various definitions and write the unitarity relations
associated with the $\pi\pi$ and the $K\pi$ two-body channels. In order to deal
with $K\pi$ we introduce two combinations of the amplitudes $W_+$,
$W_S$ which correspond to the $K\pi$ isospins $I=1/2$ and
$I=3/2$.  The $\pi\pi$ contribution is affected by the presence
of an anomalous threshold which is discussed in
sec.~\sect{2.3}. Before writing the dispersive representations, we
examine the asymptotic behaviour which is found to depend on the
$K\pi$ isospin. The dispersive representations, which are of the
Muskhelishvili-Omn\`es type~\cite{Muskhelishvili,Omnes:1958hv}, are
then written down in eqs.~\rf{W32}  and~\rf{W12nu}.
Then, in sec.~\sect{4}, the various inputs which
are needed in the evaluation of the dispersive representations, i.e. the
$\pi\pi$ and $K\pi$ form factors as well as the $K\to3\pi$ amplitudes are
discussed. These are known, except for the $\Delta{I}=1/2$ component 
of the $K_S\to\pip\pim\piz$ amplitude which has not yet been determined 
from the decay data, because it is chirally suppressed, 
but plays a role in the dispersive integrals. We will
show that, given the values of the two parameters $a_+$ and $a_S$, 
one can determine both this component of the $K_S\to\pip\pim\piz$ 
amplitude and the $W_+$, $W_S$ form factors.
The results, finally, are presented in sec.~\sect{5}.

\section{Analyticity and unitarity}\lblsect{2}

\subsection{Local and non-local contributions}
The radiative decays $K^+\to\pi^+ \ell^+\ell^-$ and 
$K_S\to\pi^0 \ell^+\ell^-$ ($l=e,\mu$) are flavour-changing
neutral-current (FCNC)  processes that are suppressed in the standard
model. They are induced by $\gamma$-penguin, $Z$-penguin and $W$-box
diagrams~\cite{Gaillard:1974hs,Okun:1975uq,Gilman:1979ud}. At low energies
(below the charm mass), these diagrams are represented by local
operators involving only the three light quarks $u,d,s$. 
They consist of a set of four-quark
operators (we use the notation of ref.~\cite{Buras:1994qa})
\be\lbl{O4q}
\mathcal{L}_{\Delta{S}=1}(x)= -\frac{G_F}{\sqrt2}V_{ud}V^*_{us}
\sum_1^6  C_i O_i(x) 
\en
and of two operators involving the direct coupling of a leptonic vector or axial current to the neutral strangeness-changing $V-A$ current, 
\be\lbl{O7V}
\mathcal{L}^{\rm lept}_{\Delta{S}=1}(x)= -\frac{G_F}{\sqrt2}V_{ud}V^*_{us}\big[
C_{7V}(\nu)O_{7V}(x)+ C_{7A}O_{7A}(x) \big]\   ,
\en
with
\be
O_{7V}= \bar{s}\gamma_\mu(1-\gamma_5){d}\, \bar{\ell}\gamma^\mu\ell  ,\quad
O_{7A}= \bar{s}\gamma_\mu(1-\gamma_5){d}\, \bar{\ell}\gamma^\mu\gamma^5\ell\ . 
\en
In the sequel we will neglect all CP-violating effects. This means that the imaginary parts of the Wilson coefficients $C_i$ and $C_{7V}$ , $C_{7A}$ are discarded and that $K_S$ will be identified with the CP-even combination of $K^0$ and ${\bar K}^0$. Moreover, we will not consider the contribution from $O_{7A}(x)$ since it plays no role in our analysis and since its contribution to the amplitude is tiny anyway.
The contributions to the radiative amplitudes associated with the
local operator $O_{7V}$ can be written in terms of the semi-leptonic $K\pi$ 
form factors $f_+^{K\pi}$, $f_-^{K\pi}$ assuming isospin symmetry,
\be\lbl{local}
\braque{\pip(p_2)\vert \bar{s}\gamma_\mu{d}\vert K^+(p_1)}=
-\braque{\piz(p_2)\vert \bar{s}\gamma_\mu{d}\vert K_S(p_1)}=
(p_1+p_2)_\mu f_+^{K\pi}(s)+ (p_1-p_2)_\mu f_-^{K\pi}(s)\ .
\en
The radiative amplitudes receive  additional non-local
contributions involving a virtual photon, which are associated 
with the matrix elements
\be
\braque{\pi\vert T\{ j_\mu^{em}(0) i \mathcal{L}_{\Delta{S}=1}(x) \}\vert{K}}
\lbl{j_L}
\en
where $j_\mu^{em}=\frac{2}{3}\bar{u}\gamma_\mu{u}-\frac{1}{3}(
\bar{d}\gamma_\mu{d}+\bar{s}\gamma_\mu{s})$ is the electromagnetic
current of three-flavour QCD. Using the same notation as in ref.~\cite{DAmbrosio:2018ytt} the two
$K\to\pi\gamma^*$ form factors are defined as follows
\be
\ba{l}
\int d^4 x 
\braque{\pi^+(p_2)\vert T \{ j_\mu^{em}(0) 
i \mathcal{L}_{\Delta{S}=1}(x) \} \vert K^+(p_1) }=
\dfrac{2W_+(s;\nu)}{16\pi^2\mkd} Q_\mu ,\\[0.3cm]
\int d^4 x 
\braque{\pi^0(p_2)\vert T \{ j_\mu^{em}(0) 
i \mathcal{L}_{\Delta{S}=1}(x) \} \vert K^0(p_1) }=
\dfrac{\sqrt2W_S(s;\nu)}{16\pi^2\mkd} Q_\mu ,\\[0.3cm]
\ea
\en
with
\be
Q_\mu=\frac{s}{2}(p_1+p_2)_\mu-\frac{\Delta_{K\pi}}{2}(p_1-p_2)_\mu   ,
\en
where $s=(p_1-p_2)^2$ is the photon virtuality and
$\Delta_{K\pi}=\mkd-\mpid$. This non-local part of the form factors depends on a renormalisation scale $\nu$~\cite{Gilman:1979ud,DAmbrosio:2018ytt}, which reflects the fact that
the time-ordered product in eq.~\rf{j_L} is singular at short distances \cite{Isidori:2005tv} and needs to be regularised.

Physical form factors (which do
not depend on the renormalisation scale) are obtained by adding 
the non-local and local contributions
\be\lbl{Wphys}
\ba{l}
{W_+(s)}={W_+(s;\nu)}
+4{\pi}\dfrac{G_F\mkd V_{ud}V_{us}^*\,}{\sqrt2\, \alpha}
C_{7V}(\nu)\,f_+^{K\pi}(s),\\[0.4cm] 
{W_S(s)}={W_S(s;\nu)}
- 4{\pi}\dfrac{G_F\mkd V_{ud}V_{us}^*\,}{\sqrt2\, \alpha}
C_{7V}(\nu)\,f_+^{K\pi}(s)\ .
\ea\en
We will construct dispersive representations of the form factors
$W_{+,S}(s)$ assuming that they satisfy the usual analyticity properties. 

\begin{figure}[ht]
\centering
\includegraphics[width=0.30\linewidth]{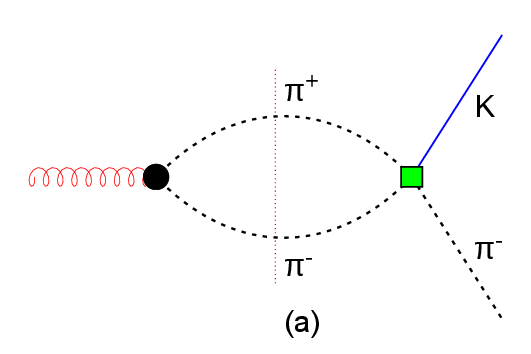}%
\includegraphics[width=0.30\linewidth]{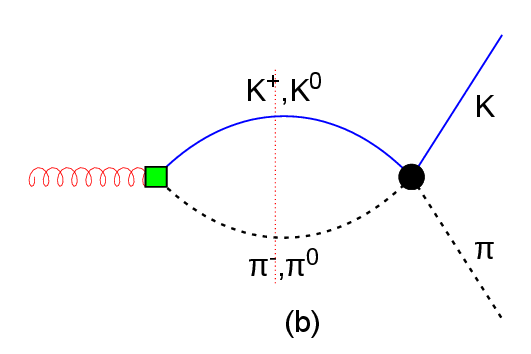}
\caption{\small Contributions to the unitarity relation from the
  $\pi\pi$ and the $K\pi$ states. A black dot represents an
  electromagnetic or strong vertex, a green square denotes a weak
  $\Delta{S}=1$ vertex}  
\label{fig:unitarity}
\end{figure}

\subsection{Unitarity: $\pi\pi$ and $K\pi$ contributions}

We discuss here the contributions from the $\pi\pi$ and $K\pi$ states in the
unitarity relations (see fig.~\fig{unitarity}) of the form factors $W_{+,S}$,
which we will be able to evaluate in a model independent way based on
experimental input. The two-pion state is the lightest hadronic state that contributes to the discontinuities of the form factors. 
In the low-energy chiral expansion this contribution starts 
at order $p^4$ and is the dominant one~\cite{Ecker:1987qi}. 
It is also important from a dispersive point of view because it 
displays a resonance peak associated with the 
$\rho(770)$ meson. The discontinuities\footnote{The discontinuity is defined
as $\disc[W(s)]\equiv (W(s+i\epsilon)-W(s-i\epsilon))/(2i)$.} 
associated with $\pi^+\pi^-$, 
illustrated in fig.~\fig{unitarity}(a), take the form
\be\lbl{discWpipi}
\ba{l}
\dfrac{\disc[W_+(s)]_{\pi\pi}}{16\pi^2\mkd}=\dfrac{\theta(s-4\mpid)}{16\pi}
\left(1-\dfrac{4\mpid}{s}\right)^{3/2} [F_V^{\pi\pi}(s)]^* \times
f_1^{\Kp\pim\to \pip\pim} (s),
\\[0.4cm]
\dfrac{\disc[W_S(s)]_{\pi\pi}}{16\pi^2\mkd}=\dfrac{\theta(s-4\mpid)}{16\pi}
\left(1-\dfrac{4\mpid}{s}\right)^{3/2} [F_V^{\pi\pi}(s)]^*\times
f_1^{K_S\piz\to \pip\pim} (s) ,
\ea\en
where $F^{\pi\pi}_V(s)$ is the electromagnetic form factor of the pion,
\be
\braque{\pip(p_2)\vert j_\mu^{em}(0)\vert \pip(p_1)}=(p_1+p_2)_\mu
F_V^{\pi\pi}((p_1-p_2)^2) ,
\en
and
$f_1^{K\pi\to\pi\pi}$ is the $J=1$ partial-wave projection of the
$K\pi\to\pi\pi$ amplitude,
\be\lbl{partialwave}
f_1^{K\pi\to\pi\pi}(s)=\frac{1}{2\kappa(s)}\int_{-1}^1 dz\, z\,
  {\cal A}_{K\pi\to\pi\pi}(s,t(z),u(z)),\
\en
with $\kappa(s)=\sqrt{ (s-m_-^2)(s-m_+^2)(1-4\mpid/s)}$,
$m_\pm=\mk\pm\mpi$.
The Mandelstam variables $s$, $t$, $u$ in the $K(p_K)\pi(p_3)\to
\pi(p_1) \pi(p_2)$ amplitudes being defined as
\be
s=(p_K+p_3)^2,\ t=(p_K-p_1)^2,\ u=(p_K-p_2)^2  ,
~ s+t+u=3s_0 , ~ s_0 \equiv \frac{\mkd}{3} + \mpid\ 
\en
the dependence of the variables $t$,$u$ on the scattering angle
$\theta$ in the $\pip\pim$ centre-of-mass system is given by
\be
t(z),u(z)=\frac{1}{2}\left(3s_0-s \pm \kappa(s)\, z\right)\ ,\
z=\cos\theta    .
\en

Next, the contributions from the $\Kp\pim$ and $\Kz\piz$ 
states to the unitarity relation are illustrated 
in fig~\fig{unitarity}(b). As can be seen from the 
figure, they are proportional to the form factors
themselves and to $K\pi\to{K}\pi$ amplitudes. 
Since $\Kz\piz$ can scatter to $\Kp\pim$ and vice versa
these unitarity contributions induce a coupling between $W_+$
and $W_S$. It is convenient, then,  
to express the dispersion relations in terms of the form 
factors which correspond to the $K\pi$
isospins $I=1/2$ and $I=3/2$ 
combinations that diagonalise 
the $S$-matrix in the $K\pi\to K\pi$ subspace,
assuming isospin conservation
\be\lbl{Wisospin}
\ba{ll}
W^{[\undemi]}(s)&\equiv  2W_+(s)-W_S(s)\\[0.2cm]
W^{[\trdemi]}(s)    &\equiv  W_+(s)+W_S(s) \ .
\ea\en
In the energy region where $K\pi\to K\pi$ scattering is elastic the
$K\pi$ unitarity relations can be  simply expressed as follows 
in terms of the $J=1$ elastic $K\pi$ phase-shifts 
$\delta_1^I$
\be\lbl{unitpiK}
\ba{ll}
\disc[W^{[\undemi]}(s)]_{{K\pi},\,{\rm elastic}}&
=\exp(-i\delta_1^\undemi(s))\sin(\delta_1^\undemi(s)) 
  \,W^{[\undemi]}(s)\\[0.2cm]
\disc[W^{[\trdemi]}(s)]_{{K\pi},\,{\rm elastic}}   &
=\exp(-i\delta_1^\trdemi(s))\sin(\delta_1^\trdemi(s)) 
  \,W^{[\trdemi]}(s)  \ .
\ea\en
In the chiral expansion, these $K\pi$ contributions start at order $p^6$
but they are important in a dispersive framework, at least in the 
 $I=1/2$ channel, where they are associated
to the peak of the $K^*(892)$ resonance. Furthermore, we 
stress that the contribution from the local operators~\rf{local} naturally drops out of the combination $W^{[\trdemi]}(s)$, and that the discontinuity equation~\rf{unitpiK} for 
$W^{[\undemi]}(s)$ holds for each component, local and 
non-local, separately, and hence for the scale-independent sum.

\begin{figure}[ht]
\centering
\includegraphics[width=0.40\linewidth]{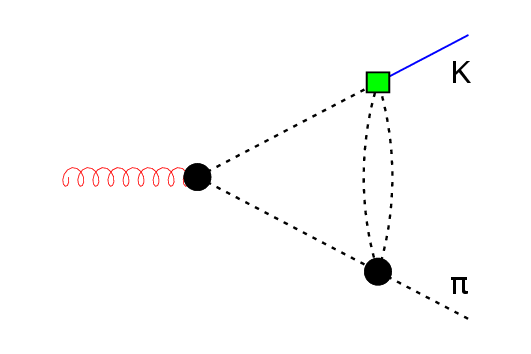}
\caption{\small Graph giving rise to an anomalous threshold}  
\label{fig:Wanomth}
\end{figure}
\subsection{Anomalous threshold}
\begin{figure}[ht]
\centering
\includegraphics[width=0.60\linewidth]{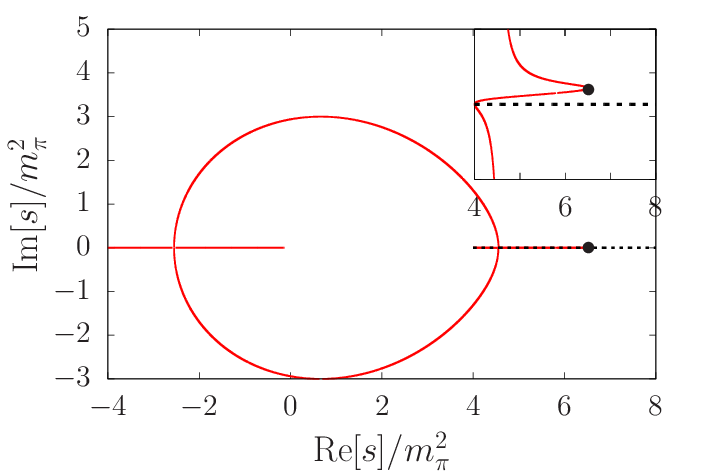}
\caption{\small Complex left-hand cut of the 
$K\pi\to \pi\pi$ partial-wave amplitudes (red line). 
The dotted line represents
the unitarity cut.
The sub-figure shows an enlarged view of the vicinity of the
  positive real axis illustrating the effect of performing an infinitesimal
  imaginary shift of $\mk$ which separates the complex cut from the
  unitarity cut. The black dot shows the position of the singularity.}  
\label{fig:cxcut}
\end{figure}

\lblsect{2.3}
It has been noticed long ago that triangle-type Feynman diagrams can exhibit
anomalous thresholds~\cite{Karplus:1958zz}. In some situations, these lead to
cuts and the associated contributions  must be added to the one from the
unitarity cut. A recent detailed discussion can be found in
ref.~\cite{Mutke:2024tww}. In the case of the form factors $W_{+,S}(s)$
considered here, using the formulas from ref.~\cite{Karplus:1958zz} one finds
that an anomalous threshold is indeed present, associated with the diagram
shown in fig.~\fig{Wanomth}. It is located at the point
$s_{\rm anom}=(\mk-\mpi)^2$ belonging to the real axis above the $\pi\pi$
threshold. In this situation, no additional cut contribution is required (the
situation is quite different in the case of, e.g., the $B\to{K}\gamma^*$ form
factors discussed in~\cite{Mutke:2024tww}). The anomalous threshold is
associated with a singularity of the form 
$(s_{\rm anom}-s)^{-3/2}$ induced in the
discontinuity functions $\disc[W_{+,S}(s)]_{\pi\pi}$ by the $J=1$ partial-wave
amplitude $f_1^{K\pi\to\pi\pi}(s)$ (see fig.~\fig{discWsum}). This singularity appears at the turning
point of the complex left-hand cut of this amplitude which is illustrated in 
fig.~\fig{cxcut}. This complex cut partly overlaps with the unitarity
$\pi\pi$ cut but it can be displaced from the real axis 
by performing an infinitesimal imaginary shift of the kaon 
mass: $\mkd \to \mkd+i\epsilon$
(e.g.~\cite{Kambor:1995yc} for a review in the case of 
$\eta\to3\pi$) such that the integral along the unitarity cut 
is well defined.

\section{A minimal set of dispersive representations}

\subsection{Asymptotic behaviour of $W^{[1/2]}(s;\nu)$ and $W^{[3/2]}(s;\nu)$}

The behaviour of the form factors $W_{+,S}(s;\nu)$ for large space-like values of
$s$ can be deduced from the operator-product
expansion of the time-ordered product \rf{j_L} that drives
the $K\to\pi\gamma^*$ transitions.  The leading term in this expansion 
is a dimension 3 operator (see refs. \cite{DAmbrosio:2018ytt} 
and \cite{DAmbrosio:2019xph} for a detailed
discussion) and the corresponding leading 
large-momentum behaviour has the following form
\be\lbl{OPEleading}
\lim_{q\to\infty}\int d^4x \, e^{iqx} T\{j_\mu^{em}(x)
  \mathcal{L}_{\Delta{S}=1}(0)\}
= (q_\mu q_\nu-g_{\mu\nu} q^2)(A+ B\ln\frac{-q^2}{\nu^2} + \cdots)\,
[\bar{s}\gamma_\nu(1-\gamma^5){d}]   
\en
where the ellipsis stands for $O(\alpha_s)$ corrections.
The operator $[\bar{s}\gamma_\nu(1-\gamma^5){d}]$ which appears in 
this expression carries isospin $I=1/2$. It thus contributes
only to the matrix element $W^{[1/2]}$ and not to 
$W^{[3/2]}$, in agreement with the absence, already 
noticed earlier, of a $\nu$-dependent contribution 
to $W^{[3/2]}$ from the local operator $O_{7V}$. 
One can verify, at lowest order, that the scale dependence 
in eq.~\rf{OPEleading} is
exactly the one needed to cancel the dependence from
$C_{7V}(\nu)$~\cite{DAmbrosio:2018ytt} in the total form factors \rf{Wphys}. 
Taking the $K\pi$ matrix element of eq.~\rf{OPEleading} one deduces that
the ratio $W^{[1/2]}(s;\nu)/f_+^{K\pi}(s)$ behaves as $\sim\log(s)$ asymptotically
and thus satisfies a convergent dispersive representation with one subtraction.

The behaviour of $W^{[3/2]} (s;\nu)$ at large space-like values of $s$ is controlled by subleading
operators in the short-distance expansion: one needs to go up to dimension-6 four-quark operators in order to find one with an isospin $I=3/2$
component, e.g.
\be\lbl{OPE_sub}
\ba{ll}
   {\displaystyle\int }d^4x\,e^{iq \cdot x}
   T\{ j^\mu(x)
     {\cal L}^{eff}_{\Delta{S=1}}(0) \}_{\rm dim=6}  \sim &
     \bigg[
C^{(6)}_V (\nu) \dfrac{g^{\mu\beta} q^\alpha -g^{\mu\alpha} q^\beta}{q^2}\\
& + C^{(6)}_A (\nu) \epsilon^{\mu\nu\alpha\beta} \dfrac{q_\nu}{q^2} \bigg]
\times(\bar{s}\gamma_\alpha{u})(\bar{u}\gamma_\beta{d}) \ {.} 
\ea
\en

%
As a consequence of this smooth behaviour at short distances, we can make the plausible assumption that the combination
$W^{[3/2]} (s)$ satisfies a dispersive representation without any subtraction.

\subsection{Muskhelishvili-Omn\`es representations}
The discontinuities of the form factors
across the $K{\pi}$ cut inside the elastic scattering region, 
given in eqs.~\rf{unitpiK}, have the form which leads to integral equations of the
Muskhelishvili-Omn\`es type \cite{Muskhelishvili,Omnes:1958hv}. These equations 
are solved in terms of the Omn\`es functions $\Omega^\undemi_{1} (s)$ and
$\Omega^\trdemi_{1} (s)$, 
\be\lbl{MOint}
\Omega^I_{1}(s)=\exp\left[ \frac{s}{\pi}\int_{m_+^2}^\infty ds'
  \dfrac{\delta_1^I(s')}{s'(s'-s) }
  \right]\ 
\en
where $m_+=\mpi+\mk$.
The $K\pi$ phases $\delta_1^I(s)$ to be used in these integrals must be 
equal to the elastic
scattering phase-shifts in the elastic region and can be chosen
in a somewhat arbitrary way above that region. It is noteworthy that the 
elastic region extends
effectively well above the first inelastic threshold $K\pi\pi$, up to
$\Lambda=1.3$ GeV approximately~\cite{Estabrooks:1977xe}.  The
$J=1$, $I=3/2$ phase-shift is very small and will be simply
interpolated to zero above 1 GeV. In the case of $I=1/2$ we can use
the $K\pi$ vector form factor (the modulus of which is a
measurable quantity) i.e. take 
$\Omega_1^{1/2}(s)\equiv f_+^{K\pi}(s)/f_+^{K\pi}(0)$ which obeys 
eq.~\rf{MOint} with a phase 
given by the $K\pi$ form factor phase in the inelastic region. 

We first consider the $I=3/2$ combination $W^{[3/2]} (s)$ and 
define the following cut-analytic function
\be
\Phi^{[3/2]}(s)=\frac{W^{[3/2]}(s)}{\Omega_{1}^{[3/2]}(s)}\ .
\en
Then, the discontinuity across the $K\pi$ cut of the Omn\`es function
in the denominator cancels exactly the discontinuity of $W^{[3/2]} (s)$ in
the numerator, such that the function $\Phi^{[3/2]}(s)$ no longer has
a $K\pi$ cut (below $\Lambda\approx1.3$ GeV). The function
$\Phi^{[3/2]} (s)$ is left with the cut associated with $\pi\pi$ firstly,
but also with unitarity cuts associated with higher-mass states like
$3\pi$, $K\bar{K},\cdots$. One remarks, however, that the
contributions from the $I=0$ resonances $\omega(782)$, $\phi(1020)$
associated with these discontinuities should be strongly reduced by
the $\Delta{I}=1/2$ rule as compared to the contribution from the
$I=1$ resonance $\rho(770)$ to $\pi\pi$. It thus seems a good
approximation to retain only the contribution from the $\pi\pi$ cut.
This leads to the following very simple representation,
\be\lbl{W32}
W^{[3/2]}(s)=
\Omega^{3/2}_{1}(s)\times\frac{1}{\pi}\int_{4\mpid}^{\Lambda^2} ds'
\frac{\disc[W_+(s')+W_S(s')]_{\pi\pi}}{(s'-s) \Omega^{3/2}_{1}(s')}   ,
\en
assuming also that the contributions from the inelastic 
region $s' > \Lambda^2$ can be neglected. The 
$\pi\pi$ discontinuities needed in
eq.~\rf{W32} were given in eq.~\rf{discWpipi}.

Considering the $I=1/2$ combination $W^{[1/2]}(s)$, in order 
to optimise the description of the contributing $\rho(770)$
and $K^*(892)$ resonances, we
start by introducing the following
once-subtracted dispersive integral over the $\pi\pi$ discontinuity
\be\lbl{W12_pipidef}
W^{[1/2]}_{\pi\pi}(s)\equiv \frac{s}{\pi}\int_{4\mpid}^{{\Lambda^2}} ds'
\frac{\disc[2W_+(s')-W_S(s')]_{\pi\pi}}{s'(s'-s) }
\en 
(which carries the information on the $\rho(770)$)
and define the following cut-analytic function
\be\lbl{Phi}
\Phi^{[1/2]}(s;\nu)=\frac{ W^{[1/2]}(s;\nu)-W^{[1/2]}_{\pi\pi}(s)
}{f_+^{K\pi}(s)} \   . 
\en
The function in the numerator no longer has a $\pi\pi$ cut 
(in the region $s<\Lambda^2$) so that 
$\Phi^{[1/2]}$ can be expressed in terms of the
$K\pi$ discontinuity (further contributions from
$3\pi$, $K\Kbar$... will be estimated approximately below). Using one
subtraction constant in the dispersive representation\footnote{
The ratio $W^{[1/2]}(s;\nu)/f_+^{K\pi}(s)$ behaving as 
$\sim\log(s)$ asymptotically, as mentioned, 
the function $\Phi^{[1/2]}(s;\nu)$ satisfies a 
once-subtracted dispersion relation provided that 
$W^{[1/2]}_{\pi\pi}(s)\sim 1/s$ when $s\to\infty$. We assume 
this to be realised through a regularisation function 
which modifies $W^{[1/2]}_{\pi\pi}(s)$ as defined 
in~\rf{W12_pipidef} when $s>>\Lambda^2$.   
}
of $\Phi^{[1/2]}(s;\nu)$ and the property that 
the ratio $W_+^{[1/2]}/f_+^{K\pi}$ 
has no $K\pi$ discontinuity in the elastic region, 
the amplitude $W^{[1/2]}(s;\nu)$ gets expressed as follows
\be\lbl{W12nu}
W^{[1/2]}(s;\nu)=  W^{[1/2]}_{\pi\pi}(s)+  f_+^{K\pi}(s)\Big[C^{[1/2]}(\nu)
-\frac{s}{\pi}\int_{m_+^2}^{\Lambda^2} ds'
\frac{W^{[1/2]}_{\pi\pi}(s')\im[1/f_+^{K\pi}(s')]}{s'(s'-s)}\Big]\  .
\en
In this representation, the dependence on the scale $\nu$ must be
entirely carried by the subtraction
constant $C^{[1/2]}(\nu)$ since all the other terms
on the right-hand side of eq.~\rf{W12nu} are manifestly scale
independent. The structure of the representation~\rf{W12nu} is the
appropriate one to allow for an exact cancellation of the 
$\nu$ dependence upon adding the contribution proportional to 
$C_{7V}(\nu)$ in eq.~\rf{Wphys}. 
The physical amplitude, after adding this part, can be written as 
\be\lbl{W12phys}
W^{[1/2]}(s)=  W^{[1/2]}_{\pi\pi}(s)+  f_+^{K\pi}(s)\Big[
G_F\mkd\frac{2a_+-a_S}{f_+^{K\pi}(0)}
-\frac{s}{\pi}\int_{m_+^2}^{\Lambda^2} ds'
  \frac{W^{[1/2]}_{\pi\pi}(s')\im[1/f_+^{K\pi}(s')]}{s'(s'-s)}\Big]\ 
\en
where $a_+$, $a_S$ are proportional to the values at $s=0$
\be
W_{+,S}(0)=G_F\mkd \,a_{+,S}\ 
\en
following the notation of the B1L model. 

At this point eqs.~\rf{W32} and \rf{W12phys} form 
a system of coupled equations for the form factors 
$W_+ (s)$ and $W_S (s)$ which exhibits an explicit  
dependence on a single free
parameter, $2a_+-a_S$. In order to make use of these equations, we next need to address the $K\pi\to\pi\pi$ partial-wave projections appearing in eq.~\rf{discWpipi}.
While our knowledge of the $K\to3\pi$ amplitudes is rather detailed,  it is, however, not fully complete. The $\Delta{I}=1/2$ part of the
$K_S\to\pip\pim\piz$ amplitude, which corresponds to the 
$3\pi$ state
with a total isospin $I_{3\pi}=0$ is chirally
suppressed~\cite{Zemach:1963bc} and has not yet been 
determined from
the available experimental data on the Dalitz plot distribution. In
the Khuri-Treiman approach of ref.~\cite{Bernard:2024ioq} this 
part is proportional to a parameter
called $\tilde{\mu}_1$, see eqs.~\rf{KS3pi} and 
\rf{tildeM1} below.
The equations~\rf{W32} and \rf{W12phys} can then be used in the following way:
given the values of $W_{+,S}$ at $s=0$, i.e.  $a_+$ and $a_S$, one can
first determine $\tilde{\mu}_1$ from eq.~\rf{W32} that provides a 
linear relation between $a_++a_S$ and $\tilde{\mu}_1$ 
(see eq.~\rf{Wsum2s=0} below).  Then, the two form
factors can be completely predicted for $s\ne0$
using the two dispersive equations. We next describe how we proceed with their evaluation.

\begin{figure}[hbt]
\includegraphics[width=0.5\linewidth]{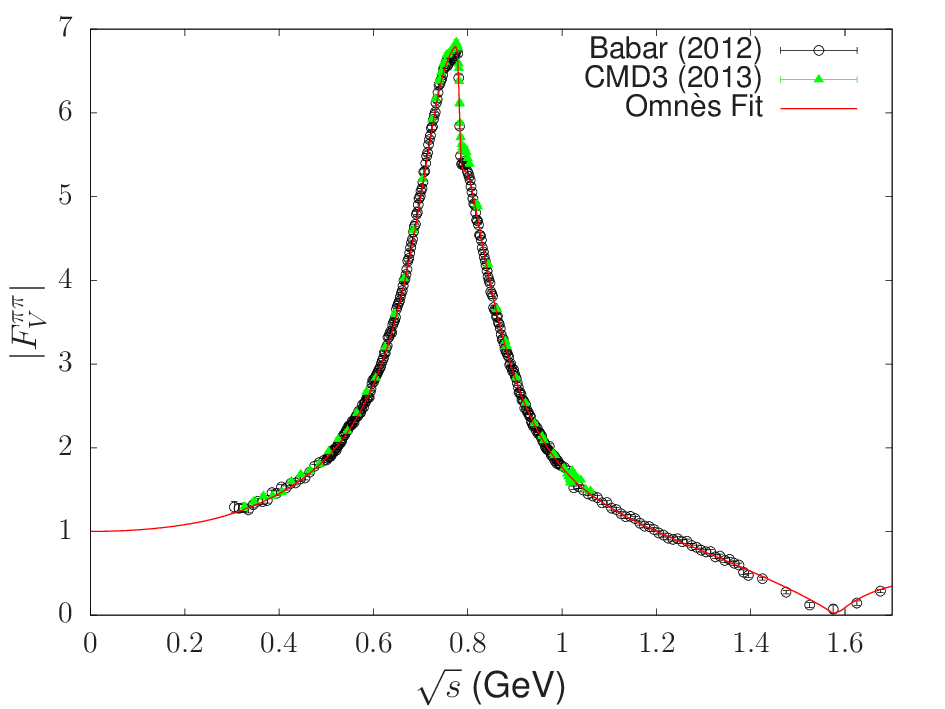}%
\includegraphics[width=0.5\linewidth]{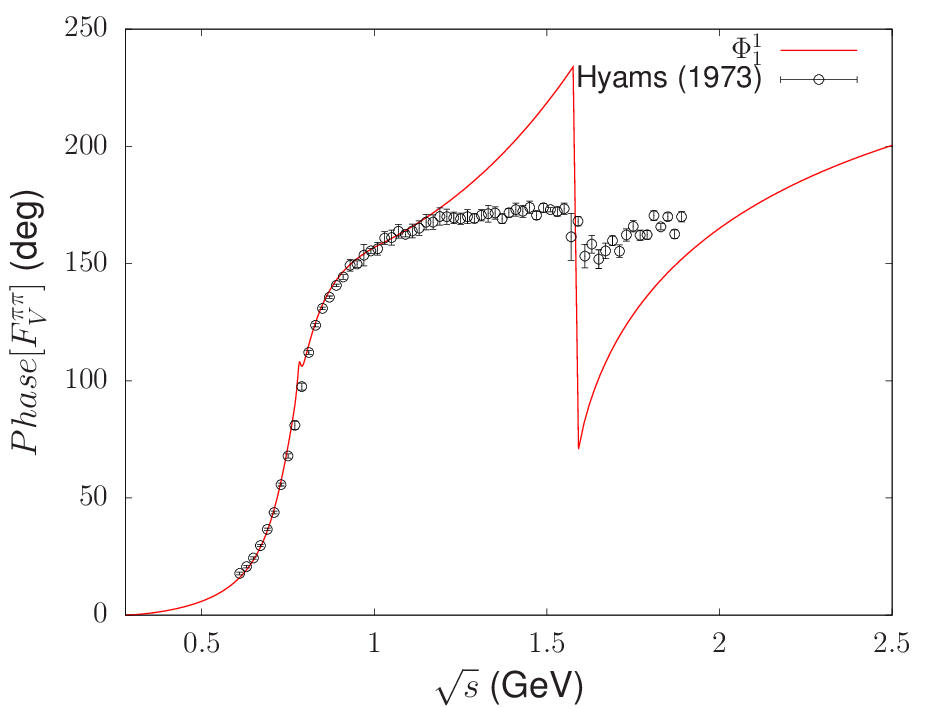}
\caption{\sl The phase of the $\pi\pi$ form factor used in the Omn\`es-type
  representation (right) and the resulting modulus (left).}
\label{fig:FVpipi} 
\end{figure}
\section{Evaluation of the dispersion relations}\lblsect{4}
\subsection{Input for the  $\pi\pi$ and $K\pi$ form factors}

We describe both the $\pi\pi$ and the $K\pi$ form factors using 
dispersive representations analogous to~\rf{MOint}, which properly encode both 
the analyticity properties and the relation between the
phase and the elastic scattering phase-shift at low energy, see e.g. 
refs.~\cite{Gourdin:1974iq,Leutwyler:2002hm}. We also assume the absence of
complex zeros~\cite{Leutwyler:2002hm} (see~\cite{Leplumey:2025kvv} for a 
recent discussion).  The phase
of the pion form factor, $\phi_1^1(s)$, is described in the elastic region
(which we take to be $s\le1.44$ $\hbox{GeV}^2$) as a sum:
$\phi_1^1(s)=\delta_1^1(s) +\delta_\omega(s)$ where $\delta_1^1$ is the
$\pi\pi$ elastic scattering phase, for which we use the experimental
results~\cite{Hyams:1973zf} with the same parameterisation as used in
the $K\to3\pi$ KT equations, and $\delta_\omega(s)$ is the isospin breaking
contribution induced by the $\omega(782)$ resonance which we parameterise as
\be
\delta_\omega(s)=\theta(s-9\mpid)\,
{\rm Arg} \left(1+\epsilon_\omega e^{i\phi_\omega}\frac{m_\omega^2}
{(m_\omega-i\Gamma_\omega/2)^2-s}\right)\ .
\en
In the inelastic region we satisfy ourselves with a rough description
(see ref.~\cite{RuizArriola:2024gwb} for a recent detailed determination of
this phase), since we need the form factor only in the elastic
region. The free parameters in the phase are fitted to experimental
data on the modulus of the form factor. Because of the importance of
$F_V^{\pi\pi}$ in the evaluation of the $g-2$ of the muon, there has
been a number of measurements of this quantity in recent years
using $e^+e^-\to \pip\pim$ scattering as well as $\tau^\mp\to \pi^\mp\piz\nu$
decays, see the review \cite{Davier:2023fpl}. More specifically, we
have used the data from Babar~\cite{Lees:2012cj} in the present
work. The results of our fit for the modulus and the phase of
$F_V^{\pi\pi}$ are shown in fig.~\fig{FVpipi}.

\begin{figure}[htb]
\includegraphics[width=0.5\linewidth]{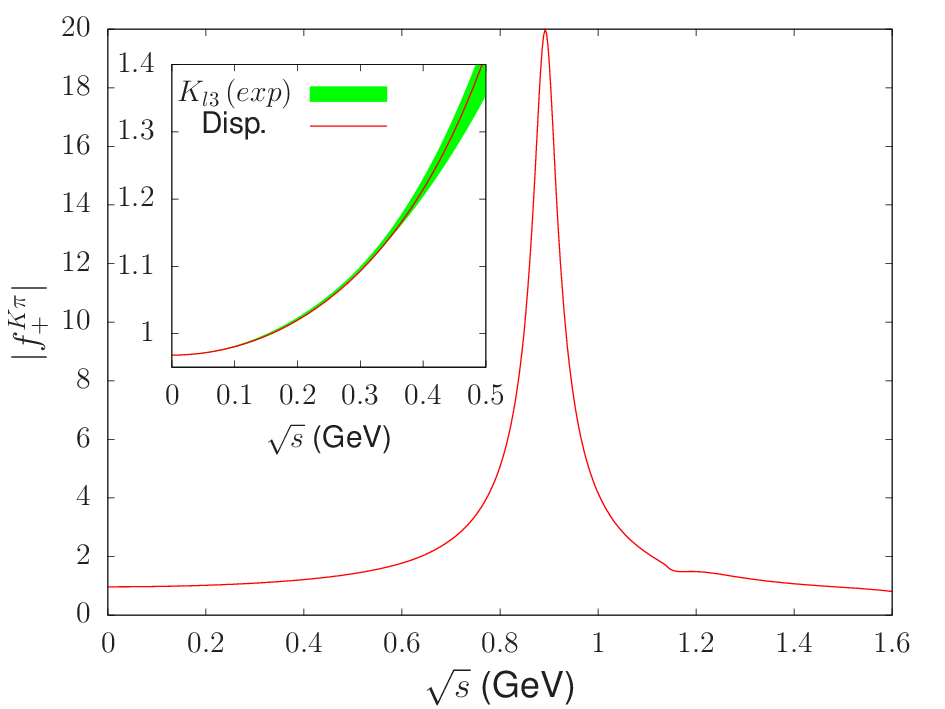}%
\includegraphics[width=0.5\linewidth]{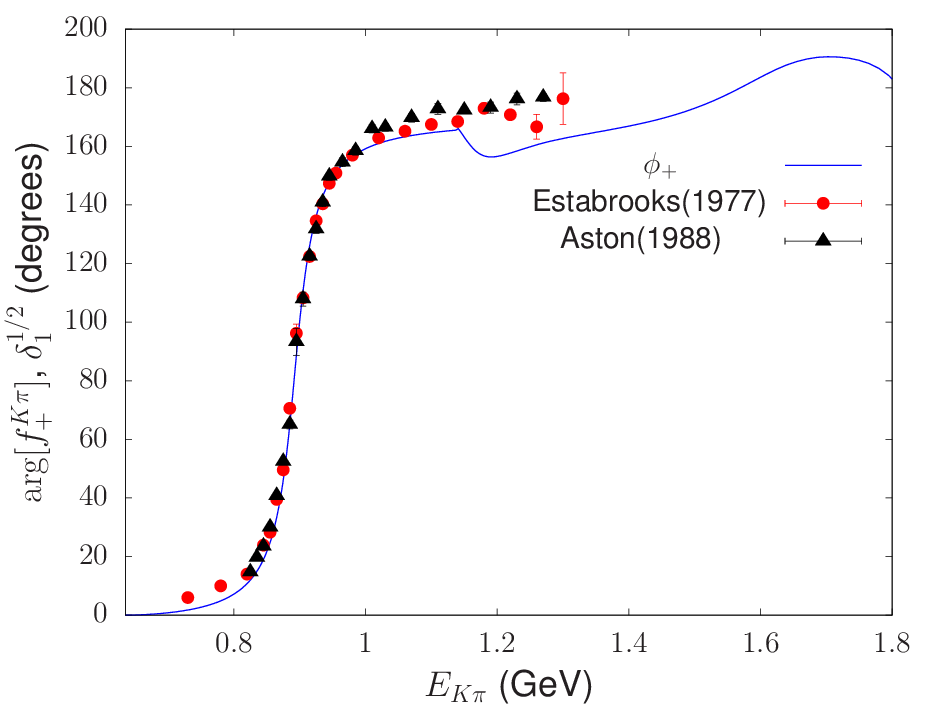}
\caption{\sl Phase (right) and corresponding modulus (left) of the $K\pi$ vector form
factor $f_+^{K\pi}(s)$. The phase is compared with the $K\pi$ $P$-wave scattering
phase-shift from refs. \cite{Estabrooks:1977xe,Aston:1987ir}. The modulus is
compared at low energy to a quadratic representation of experimental $K_{l3}$ 
decay data\cite{KLOE:2006kms}.} 
\label{fig:fplus} 
\end{figure}

The $K\pi$ vector form factor can be determined, in principle, from
$\tau\to K\pi\nu$ decays and, at low energy, from $K_{l3}$ decays. The
corresponding amplitudes involve both $f_+^{K\pi}$ and the scalar form factor
$f_0^{K\pi}$,
\be
f_0^{K\pi}(s)\equiv f_+^{K\pi}(s)+\frac{s}{\Delta_{K\pi}} f_-^{K\pi}(s)\ .
\en
Several theoretical models of these two form factors have been proposed 
(e.g.~\cite{Beldjoudi:1994hi,Moussallam:2007qc,Jamin:2008qg,Boito:2010me,Bernard:2013jxa}) 
while measurements of $\tau\to K\pi\nu$
decays have been performed by Aleph~\cite{ALEPH:1999uux},
Belle~\cite{Epifanov:2007rf} and Babar~\cite{Adametz:2011zz}.  We use
here a two-channel unitary model for the scalar form factor as originally proposed
in~\cite{Jamin:2001zq}. For the vector form factor we use a 
dispersive phase representation with a normalising value at $s=0$,
$f_+^{K\pi}(0)=0.968$, and in which the phase $\phi_+(s)$ is constructed as
follows: a) in the elastic scattering region ($\sqrt{s} \le1.14$ GeV)
it is described as a simple Breit-Wigner function with two parameters
$M_{K^*}$, $\Gamma_{K^*}$
\be
\phi_+(s)=\arctan\left(\frac{\sqrt{s}\Gamma_{K^*}(s)}{M^2_{K^*}-s}\right),\quad
\Gamma_{K^*}(s)=\Gamma_{K^*} \frac{M^2_{K^*}}{s}\left(
\frac{p_{cm}(s,\mpi,\mk)}{p_{cm}(M^2_{K^*},\mpi,\mk)}
\right)^{3/2}
\en
($p_{cm}$ being the centre-of-mass momentum) and b) in the inelastic region the phase $\phi_+(s)$ is taken from the
three-channel unitary model proposed in
ref.~\cite{Moussallam:2007qc}. While model dependent, this phase has
the nice property of going to $\pi$ asymptotically
from above, as required by QCD~\cite{Leutwyler:2002hm}, and its energy
dependence reflects the presence of the $K^*(1410)$ and $K^*(1680)$
resonances. A rather good fit of the $\tau\to{K}\pi\nu_\tau$ Belle 
data~\cite{Epifanov:2007rf} can be obtained in this manner with 
\be
M_{K^*}=(895.51\pm 0.16)\, \hbox{MeV},\
\Gamma_{K^*}=(48.25\pm 0.35)\, \hbox{MeV}
\en
and a $\chi^2$ value  $\chi^2/N_{dof}=132/98$. Fig.~\fig{fplus} (right) shows
the phase $\phi_+$ and illustrates its compatibility with the elastic
$P$-wave scattering phase-shifts
from refs. \cite{Estabrooks:1977xe,Aston:1987ir} in the elastic energy
region. The left plot shows the modulus
$\vert{f^{K\pi}_+}(s)\vert$ computed from the dispersive phase
representation. At low energy it is seen to be in very good agreement with the experimental
determination from $K_{l3}$ decays as given in
ref.~\cite{KLOE:2006kms}.

\begin{figure}[htb]
  \centering
\includegraphics[width=0.5\linewidth]{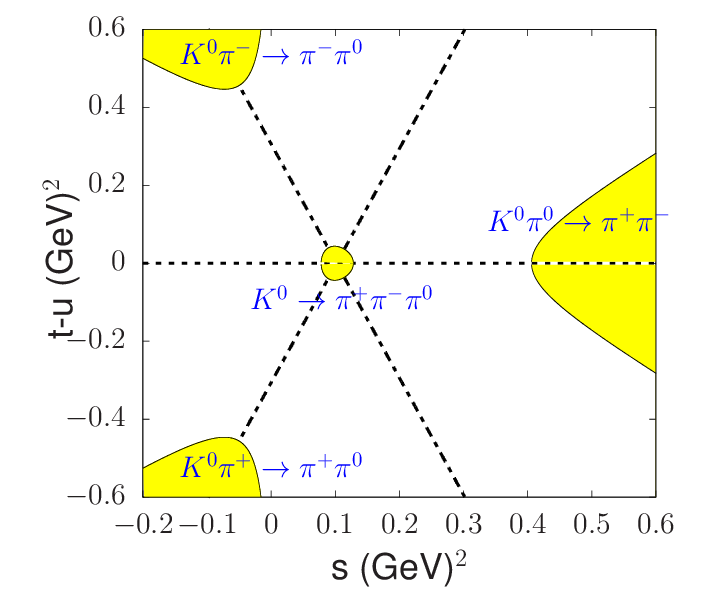}
\caption{\sl Illustration of the various physical regions for $K^0\to3\pi$ and
$K^0\pi\to\pi\pi$ as a function of the Mandelstam variables $s$ and
  $t-u$. Also shown are the three lines $t=u$, $s=t$,$u=s$.}
\label{fig:Mandelstam} 
\end{figure}

\subsection{Input for the $K\pi\to \pi\pi$ amplitudes}

In order to compute the discontinuities displayed in eq.~\rf{discWpipi}
we need the amplitudes $K\pi\to\pi\pi$ in an energy range extending from the
$\pi\pi$ threshold up to physical scattering region in order to capture the
effects of the $\rho(770)$ resonance. 
For this purpose, we rely on a set of
solutions of the Khuri-Treiman
equations \cite{Khuri:1960zz,Sawyer:1960hrc} recently obtained in 
ref.~\cite{Bernard:2024ioq}. These equations, by
construction, encode the properties of analyticity and elastic unitarity 
(for the $J=0,1$ partial-waves) that allow
one to perform the extrapolation from the Dalitz plot region, in which
$K\to3\pi$ decay measurements can be performed, up to the physical scattering
region. These various regions of the Mandelstam plane are illustrated in 
fig.~\fig{Mandelstam}.

In the KT formalism as used in~\cite{Bernard:2024ioq},
the $K\pi\to\pi\pi$ amplitudes are expressed in terms of a set of nine
functions of one variable. Four of these ($M_0$, $M_1$, $M_2$, $\widetilde{M}_1$) describe $\Delta{I}=1/2$
transitions and the remaining
five ($N_0$, $N_1$, $N_2$, $\widetilde{N}_1$, $\widetilde{N}_2$)
describe $\Delta{I}=3/2$ transitions. We reproduce
below the expressions of the two amplitudes which are of interest here
in terms of these functions
\be
\ba{ll}
{\cal A}_{\Kp\pim\to \pip\pim}(s,t,u)= & \Big\{(t-u)\left(M_1(s)+N_1(s)
+\frac{3}{2}\widetilde{N}_1(s)\right) \\[0.2cm]
\ &+M_0(t)+N_0(t)+\frac{1}{3}\left(M_2(t)+N_2(t)\right)
-\frac{1}{2}\widetilde{N}_2(t) \\[0.2cm]
\ & +(t\leftrightarrow{s})\Big\} +2\left(M_2(u)+N_2(u)\right)+\widetilde{N}_2(u)
\ea\en
and
\be\lbl{KS3pi}
\ba{ll}
 {\cal A}_{K_S\piz\to \pip\pim}(s,t,u)= & (t-u)\widetilde{M}_1(s)
+(s-t)\widetilde{M}_1(u)+(u-s)\widetilde{M}_1(t) \\[0.2cm]  
\ & -2(t-u)\widetilde{N}_1(s)
+\Big\{ \widetilde{N}_2(t)+(s-t)\widetilde{N}_1(u)
-(t\leftrightarrow{u})\Big\}\ .
\ea\en
Performing the angular integrations of these amplitudes according to
eq.~\rf{partialwave} one obtains the two $J=1$ partial waves in the following
form,
\be\lbl{dzzintegrals}
\ba{ll}
f_1^{\Kp\pim\to    \pip\pim}(s)
& = \dfrac{1}{3}\big(M_1(s)+N_1(s)+\frac{3}{2}
  \widetilde{N}_1(s) 
   +\widehat{M}_1(s)+\widehat{N}_1(s)+
\frac{3}{2}\widehat{\widetilde{N}}_1(s)\big)\\[0.3cm]
f_1^{K_S\piz\to\pip\pim}(s)
& = \dfrac{1}{3}\big(\widetilde{M}_1(s)+\widehat{\widetilde{M}}_1(s)
\big) 
   -\dfrac{2}{3}\big( \widetilde{N}_1(s) +     \widehat{\widetilde{N}}_1(s)\big)\ .
   \ea\en
The hat-functions, $\widehat{M}_1$, $\widehat{N}_1,\,\cdots$
(following the notation introduced in~\cite{Anisovich:1996tx}) 
in eq.~\rf{dzzintegrals}
represent linear combinations of angular integrals of the
one-variable functions (explicit expressions can be found
in ref.~\cite{Bernard:2024ioq}). The KT equations ensure that the
partial waves $f_1^{K\pi\to\pi\pi}$ from eqs~\rf{dzzintegrals} 
satisfy the elastic $\pi\pi$ unitarity relations. They should thus be 
valid at low energy and up to the $\rho(770)$ mass region, but no longer
at higher energy in particular in the $K^*(892)$ mass region since
$K\pi$ unitarity is not implemented.

The nine one-variable functions 
can be expressed linearly in terms of a set of
independent solutions of the Khuri-Treiman integral 
equations. These linear representations involve 
11 subtraction constants $\mu_k, \nu_k$, $k=0,1,2,3$, 
$\tilde{\nu}_0$, $\tilde{\nu}_1$, and $\tilde{\mu}_1$. For
instance, for $M_1$, $N_1$, $\tilde{N}_1$ one has
\be
M_1(s)=\sum_{k=0}^3 \mu_k\, {\cal S}_1^{(k)}(s),\
N_1(s)=\sum_{k=0}^3 \nu_k\, {\cal S}_1^{(k)}(s),\
\widetilde{N}_1(s)=\sum_{k=0}^1 \tilde{\nu}_k \widetilde{{\cal S}}_1^{(k)}(s),\
\en
where ${\cal S}_1^{(k)}$, $\widetilde{{\cal S}}_1^{(k)}$ are independent solutions.
In the case of $\widetilde{M}_1$, a single parameter is involved and one has
\be\lbl{tildeM1}
\widetilde{M}_1(s)=\tilde{\mu}_1 \widetilde{{\cal S}}_1(s)\ ,\quad 
\widehat{\widetilde{M}}_1(s)=\frac{\tilde{\mu}_1}{2\kappa(s)}
\int_{-1}^{+1} dz(-9(s-s_0)z-3\kappa(s)z^2)\widetilde{S}_1(t(z))\ 
\en
with $s_0=\mkd+\mpid/3$. 
The 10 coefficients $\mu_k$, $\nu_k$, $\tilde{\nu}_k$ have been determined in
ref.~\cite{Bernard:2024ioq} using experimental data on the Dalitz plot
behaviour of the $K\to3\pi$ decays.
The $11^{th}$ coefficient, $\tilde{\mu}_1$, appears, as already mentioned, only in the 
$K_S\to \pip\pim\piz$ amplitude and could not be
determined from the decay data. The reason for this can be clearly seen from eq.~\rf{KS3pi}:
the part involving the function $\widetilde{M}_1$ (first line in this
equation) is antisymmetric with respect to the three permutations
$t\leftrightarrow{u}$, $u\leftrightarrow{s}$,
$s\leftrightarrow{t}$ and thus vanishes along the three lines $t=u$, $u=s$, $s=t$
which meet at the centre of the Dalitz plot.
This part of the $K_S\to\pip\pim\piz$ amplitude, even though enhanced
in principle by the $\Delta{I}=1/2$ rule, is thus strongly suppressed
with respect to the $\Delta{I}=3/2$ part, which is antisymmetric under
$t\leftrightarrow{u}$ exchange only, in the whole physical decay region.
Here, however, the $K_S\to \pip\pim\piz$ amplitude will be used away from the 
physical decay region. In this situation
the $\Delta{I}=1/2$ part of the amplitude is no longer suppressed, it should be
comparable to or even larger than the $\Delta{I}=3/2$ amplitude. 
This feature will allow us to obtain a determination of 
the parameter $\tilde{\mu}_1$.

\begin{figure}[ht]
\centering
\includegraphics[width=0.50\linewidth]{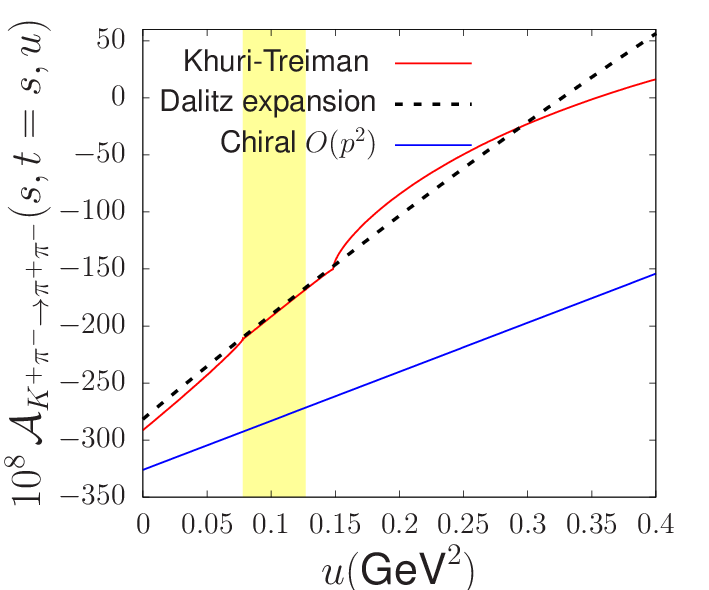}%
\includegraphics[width=0.50\linewidth]{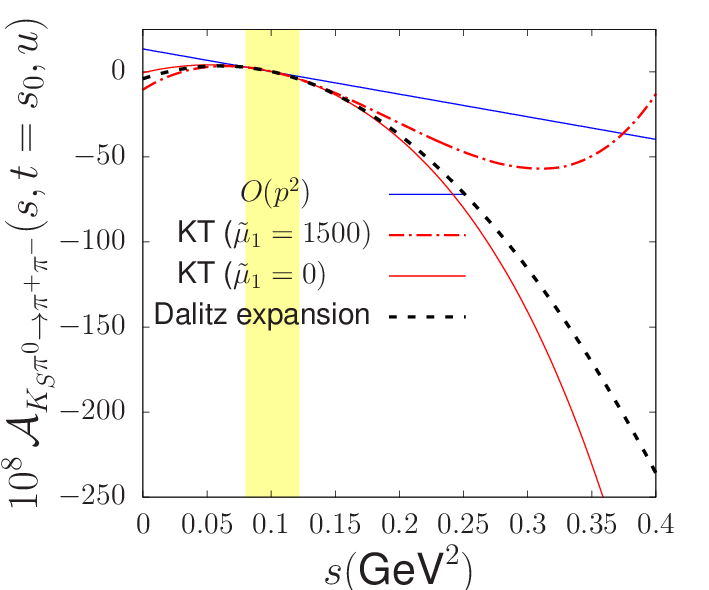}
\caption{\small Comparison of the real parts of the 
Khuri-Treiman amplitudes
with their Dalitz-plot expansions at quadratic order. The yellow 
band indicates the physical region.
The left figure shows the
$K^+\pi^-\to\pi^+\pi^-$ amplitude along the line $s=t$ as a function
of the variable
$u$. The figure on the right shows the 
$K_S\pi^0\to\pi^+\pi^-$ amplitude
fixing $t=s_0$ as a function of the variable $s$. The figures also 
show the chiral expansions of both amplitudes at order $p^2$, 
taking
$G_8=G_F/\sqrt2 V_{ud}V_{us}\times3.62$ and
$G_{27}=G_F/\sqrt2 V_{ud}V_{us}\times0.29$ from the 
review \cite{Cirigliano:2011ny}. }
\label{fig:signeG8}
\end{figure}

\section{Results}\lblsect{5}

\subsection{$K\to3\pi$: Dispersive amplitudes versus their $X$,$Y$ expansion}

Because of the importance of the $K\to3\pi$ amplitudes 
in the construction of the $W_{+,S}$ form factors let us
start with a brief comparison between the amplitudes derived from 
the Khuri-Treiman formalism and their expansions to quadratic 
order as a function of the Dalitz variables $X$,$Y$ as used in 
the B1L description~\cite{DAmbrosio:1998gur}. 
This is illustrated in fig.~\fig{signeG8}.  
The figure also shows the $K\to3\pi$ amplitudes computed 
at chiral order $p^2$ (entering in the evaluation of
$W_{+,S}$ at chiral order $p^4$). We recall their expressions
below (e.g.~\cite{Bijnens:2002vr})
\be\lbl{K3piOp2}
\ba{ll}
\left.\mathcal{A}_{\Kp\pim\to\pip\pim}(s,t,u)\right\vert_{p^2}& =G_8(u-\mkd-\mpid)
+G_{27}(-\frac{13}{3}u+\mkd+\frac{13}{3}\mpid)\\
\left.\mathcal{A}_{K_S\piz\to\pip\pim}(s,t,u)\right\vert_{p^2}& =G_{27}(t-u)\dfrac{-15\mkd+10\mpid}
{6(\mkd-\mpid)}\ .
\ea
\en
We first remark that the signs of the real parts of the 
$K\to3\pi$ KT amplitudes are in agreement with those used 
in the B1L model~\cite{DAmbrosio:1998gur} and 
they also agree with the chiral $O(p^2)$ amplitudes provided 
the two chiral couplings $G_8$ and $G_{27}$ are taken to be positive.
These signs will 
enable us to define the signs and the phases of the $W_+$, $W_S$ amplitudes.
Figure~\fig{signeG8} also shows that the Dalitz 
expansion provides a precise description\footnote{We used the 
expansion coefficients as given in Table 5 of 
ref.\cite{Bernard:2024ioq}. Let us note that the real part of the parameter
$\xi'_3$ was given with the wrong sign in this table. }
inside the physical decay regions (yellow bands in the
figures) that remains qualitatively acceptable in a 
rather large region.
Figure~\fig{signeG8} (right) shows the effect of the 
parameter $\tilde{\mu}_1$ which induces a cubic term in 
the Dalitz expansion. This term becomes important away from the
physical decay region.

\subsection{Determination of $\tilde{\mu}_1$}

\begin{figure}[bt]
\centering
\includegraphics[width=0.55\linewidth]{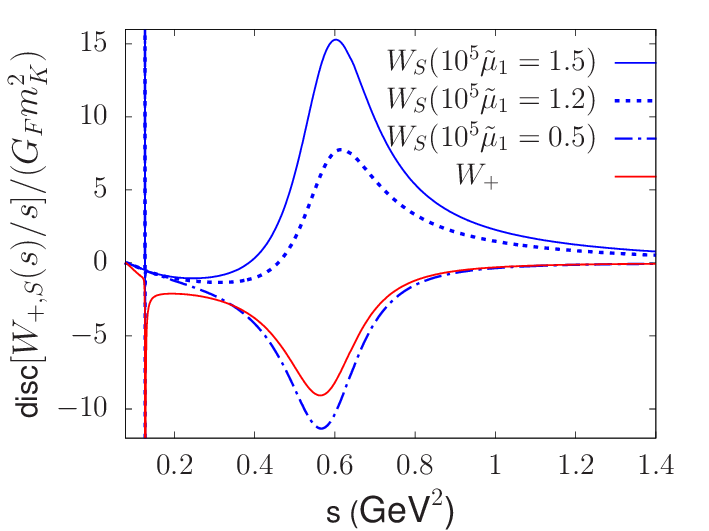}
\caption{\sl Illustration of the behaviour of the 
$\pi\pi$ discontinuities $\disc[W_+]_{\pi\pi}$ and
$\disc[W_S]_{\pi\pi}$ as a function of $s$. Notice the 
singularity at $s=(\mk-\mpi)^2$ (see sec.~\sect{2.3}). 
The function $\disc[W_S]_{\pi\pi}$ is shown for several values 
of the parameter $\tilde{\mu}_1$.}
\label{fig:discWsum}
\end{figure}

We have argued that the combination 
$W^{[3/2]} (s) = W_+(s)+W_S(s)$ that corresponds
to a $K\pi$ isospin $I=3/2$ should satisfy a dispersive 
representation with no subtraction parameter. We also assumed 
that the $\Delta{I}=1/2$ rule leads to a suppression of the 
contributions from the $I=0$ resonances relative to 
those of the $I=1$ resonances in
this combination.  Then, for small values of $s$, one should be able
to compute $W_+(s)+W_S(s)$ reliably by including the dominant
contribution from the $\rho(770)$ resonance. 
In the dispersive approach, this contribution is included in the 
$\pi\pi$ discontinuity through both the pion form
factor $F^{\pi\pi}_V$ and the $K\pi\to\pi\pi$ amplitudes. 
These amplitudes are known up to a linear dependence on
the parameter $\tilde{\mu}_1$ appearing in the $K_S\to\pip\pim\piz$
amplitude, see eqs. \rf{dzzintegrals} and \rf{tildeM1}.
Fig.~\fig{discWsum} illustrates the influence of this parameter 
on the discontinuity function $\disc[W_S]_{\pi\pi}$. 
One sees that the dependence is small at low energies $\sqrt{s}\lapprox 0.5$
GeV, as expected, but  significant in the region of the 
resonance peak.
Computing the dispersive integral for $W^{[3/2]}$ from 
eq.~\rf{W32} at $s=0$ leads to the following linear 
relation between the parameter $\tilde{\mu}_1$ and $a_++a_S$
\be\lbl{Wsum2s=0}
\ba{ll}
10^8\,G_F\mkd\,(a_+ + a_S)= & -952.4\pm30.7+i(2.17\pm0.82) \\
 & +10^8\,\tilde{\mu}_1\,(0.73\pm0.04+i(1.12\pm1.39)\cdot10^{-3})\ ,
\ea\en
where the central value corresponds to a cutoff $\Lambda=1.2$ GeV 
and the errors  are generated by varying the Khuri-Treiman 
parameters and increasing the value of the integration cutoff up
to 1.8 GeV.

Making use of the relation~\rf{Wsum2s=0}, the pair of dispersive 
representations~\rf{W32},~\rf{W12phys} provide a determination of $W_+$ and $W_S$
in terms of the two parameters $a_+$, $a_S$.
The value of $a_+$ has been determined by a number of experiments by
measuring the energy distribution in $K^\pm\to \pi^\pm e^+ e^-$ or
$K^\pm\to \pi^\pm \mu^+ \mu^-$ and extrapolating to $s=0$ using the
B1L model or simpler descriptions (see ref.~\cite{DAmbrosio:2018ytt}
for a detailed discussion and a list of references). Let us quote here
the recent results by the NA62 collaboration~\cite{NA62:2022qes}:
\be\lbl{a+NA62}
\ba{l}
a_+=-(0.575\pm0.012)\ ,\quad \chi^2/ndf=45.1/48 \\ 
a_+=+(0.373\pm0.012)\ ,\quad \chi^2/ndf=56.4/48
\ea\en
based on the $\Kp\to \pip \mu^+\mu^-$ channel. 
Estimates of $a_S$ have been derived from the measurements of the $K_S\to
\pi^0\ell^+\ell^-$ branching fractions using the B1L model additionally
assuming a VMD-type relation between $b_S$ and $a_S$. 
From the $e^+e^-$ channel, ref.~\cite{NA481:2003cfm}
gives $a_S=\pm( 1.06^{+0.26}_{-0.21} \pm0.07 )$ while in ref.~\cite{Buchalla:2003sj}
the value $a_S=\pm( 1.08^{+0.26}_{-0.21})$ is found.
From the $\mu^+\mu^-$ channel, ref.~\cite{NA481:2004nbc} obtains 
$a_S=\pm( 1.54^{+0.40}_{-0.32} \pm0.06)$. A value of $|a_S|$ lying in the range 
$[1,1.5]$ has also been obtained in refs.~\cite{Friot:2004yr,DAmbrosio:2024ncc}. 
The sum $a_+ + a_S$ should therefore satisfy 
\be
-2 \lapprox \left.a_+ + a_S\right\vert_{exp} \lapprox +2\ . 
\en
Using eq.~\rf{Wsum2s=0} we deduce that $\tilde{\mu}_1$ must be 
positive and lie in the range 
\be\lbl{mu1range}
519 \lapprox 10^8 \tilde{\mu}_1 \lapprox 2089\ .
\en
The order of magnitude of $\tilde{\mu}_1$ is reasonable if 
we compare it
with the parameter $\mu_3$ which plays an analogous role in the
description of the $\Delta{I}=1/2$ function $M_1$ and 
was determined
to be $10^8\mu_3=-1072.6\pm13.6$ in 
ref.~\cite{Bernard:2024ioq}. The
analogous parameters  that appeared in the $\Delta{I}=3/2$ functions
$N_1$, $\tilde{N}_1$ were somewhat smaller $10^8\nu_3=123.9\pm12.6$
and $10^8\tilde{\nu}_1=433.2\pm12.2$ in agreement with the expectation
from the $\Delta{I}=1/2$ enhancement rule. This rule would lead to
favour the upper half of the range~\rf{mu1range} and, then, to the
expectation that $a_+ + a_S$ should be positive.

\begin{figure}[htb]
\centering  
\includegraphics[width=0.6\linewidth]{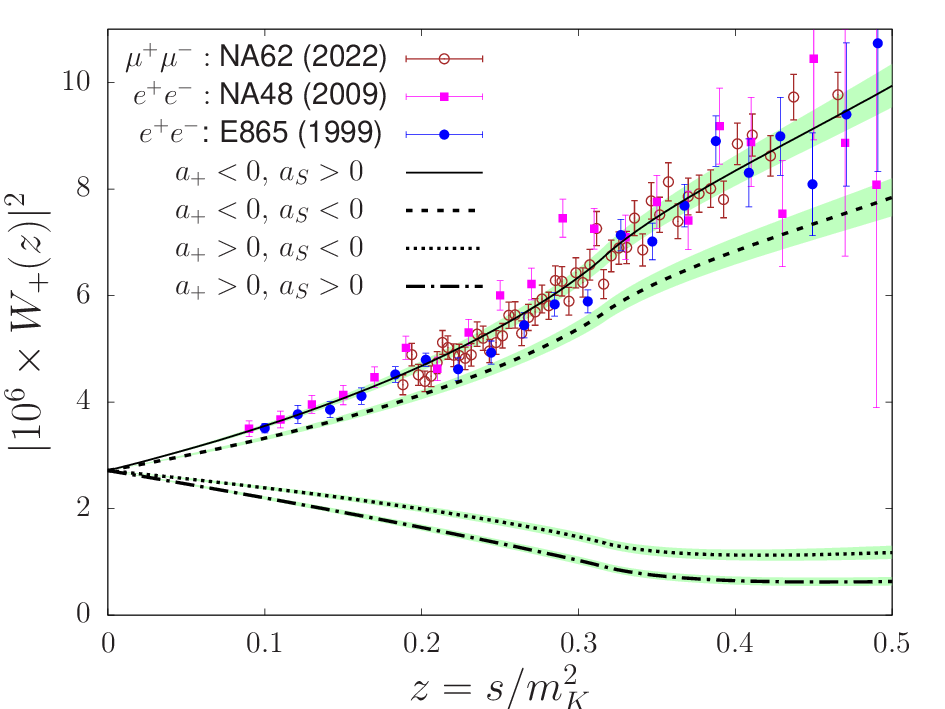}
\caption{\sl  Results for $|W_+|^2$ corresponding to different choices of
  signs for $a_{+,S}$: $a_+=\mp 0.575$, $a_S=\pm1.06$, compared to the experimental data from
  refs.~\cite{Appel:1999yq,Batley:2009aa,NA62:2022qes}}
\label{fig:W2_plus}
\end{figure}
\subsection{Results for $W_+(s)$}

Let us now present the results for the amplitude $W_+$. 
Being deduced from a coupled system of equations it 
is expected to depend on both $a_+$ and $a_S$. 
For illustration, we choose $a_+=\mp{0.575}$ (the negative value corresponding 
to the central result from the NA62 experiment~\cite{NA62:2022qes}) 
and $a_S=\pm{1.06}$ (from ref.~\cite{NA481:2003cfm}). 
The results for $|W_+ (s)|^2$ corresponding to
the four combinations of $a_+$, $a_S$ sign values are shown in fig.~\fig{W2_plus} 
and compared with the experimental determinations from
refs. \cite{Appel:1999yq,Batley:2009aa,NA62:2022qes}. The figure shows
that a positive value for $a_+$ can be ruled out
since, in this case, $|W_+|^2$ is a decreasing function 
of $s$ (see also below),
which is in disagreement with the 
experimental results. The figure also shows that 
$W_+$ has a clear
sensitivity to the value of $a_S$. The best agreement with the
experimental data is achieved when $a_S$ is positive. 
The error bands which are shown in the figure are generated by 
varying the Khuri-Treiman parameters as well as the 
integration cutoff parameter $\Lambda$. 
An additional uncertainty, however, is associated with 
the contributions
of the $I=0$ resonances $\omega(782)$, $\phi(1020)$ 
which we will 
discuss below. The pair of values $a_+=-0.575$, 
$a_S=+1.06$ leads to an amplitude $|W_+|$ which 
agrees reasonably well with experiment.
Computing the corresponding branching fractions we find
\be\lbl{K+BFs}
\ba{l}
BF[K^+\to\pip e^+e^-]=(3.01\pm0.05)\cdot10^{-7}\\ 
BF[K^+\to\pip \mu^+\mu^-]=(8.89\pm0.24)\cdot10^{-8} \
\ea\en
which are consistent with the experimental values 
(from the PDG~\cite{ParticleDataGroup:2024cfk})
\be\lbl{K+BFexp}
\ba{l}
BF[K^+\to\pip e^+e^-]_{exp}=(3.00\pm0.09)\cdot10^{-7}\\
BF[K^+\to\pip \mu^+\mu^-]_{exp}=(9.17\pm0.14)\cdot10^{-8} \
\ea\en
For small values of $s$ a linear approximation of the real
part of $W_+ (s)$ holds $W_+(s)\simeq G_{\rm F} [
a_+ \mkd + \bar{b}_+ s ]$ in which the slope parameter 
$\bar{b}_+$ can be shown from the dispersive representations
to be a linear function of $a_+$ and $a_S$ of the 
following form
\be \lbl{W_+linear}
\bar{b}_+ = -0.551 + 0.214\,a_+ - 0.100\,a_S  \ . 
\en
This formula shows that the slope of the function
$W_+ (s)$ at the origin is always negative provided 
$a_+$, $a_S$ have the
usual order of magnitude. This explains why the modulus 
of  $W_+ (s)$ is a decreasing function of $s$ when 
$a_+$ is positive, see fig.~\fig{W2_plus}.

\begin{figure}[hbt]
\centering  
\includegraphics[width=0.6\linewidth]{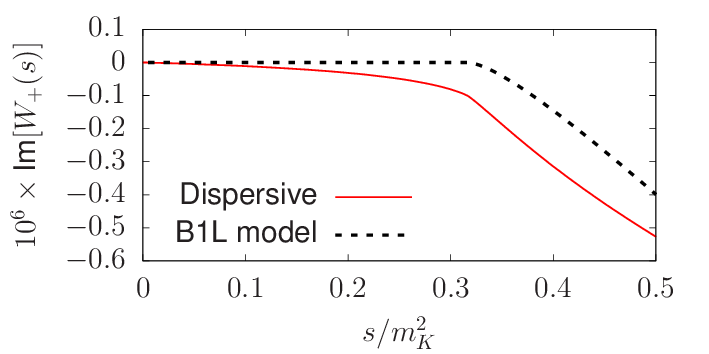}
\includegraphics[width=0.6\linewidth]{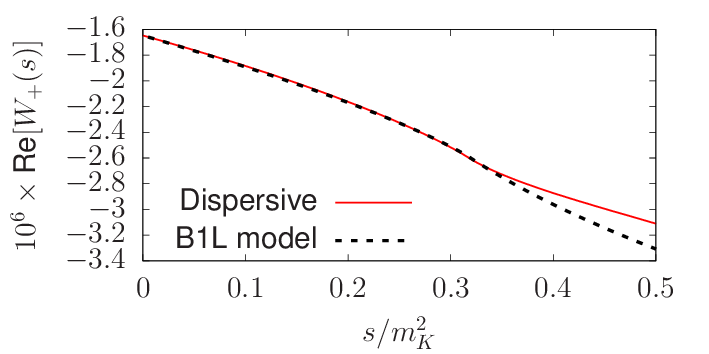}
\caption{\sl  Comparison of the real (lower plot) and imaginary parts
  (upper plot) of the $W_+$ amplitude computed from the dispersive
  representation with $a_+=-0.575$, $a_S=1.06$ (red lines) 
  and from the B1L one with the same $a_+$ and $b_+=-0.722$ (black dashed lines).}
\label{fig:compdispB1L}
\end{figure}

Finally, it is interesting to compare the amplitude $W_+$ computed from
the pair of dispersive representations~\rf{W32}~\rf{W12phys} to the one
evaluated with the B1L expression~\rf{B1L} using the 
central values
of the parameters determined by the NA62 experiment\footnote{
We note that the parameter $b_+$ in the B1L model is numerically very close
to the parameter $\bar{b}_+$ corresponding to the derivative of $W_+$ at s=0,
$\bar{b}_+=b_+-0.092$.
}:
$a_+=-0.575$, $b_+=-0.722$. This 
is shown in fig~\fig{compdispB1L}. The real parts are seen to be nearly
identical below the $\pi\pi$ threshold and slightly different above. The
imaginary parts differ somewhat more significantly: the dispersive
part being  larger in magnitude than the B1L one by approximately a factor of two in the upper part of the physical energy range. One also notices that the
imaginary part in the dispersive amplitude does not vanish below the $\pi\pi$
threshold. This reflects a contribution to the imaginary part associated with
$K\to3\pi$ followed by $3\pi\to\gamma^*\pi$. In the chiral expansion this
contribution appears at two-loops. 

\begin{figure}[htb]
\centering  
\includegraphics[width=0.5\linewidth]{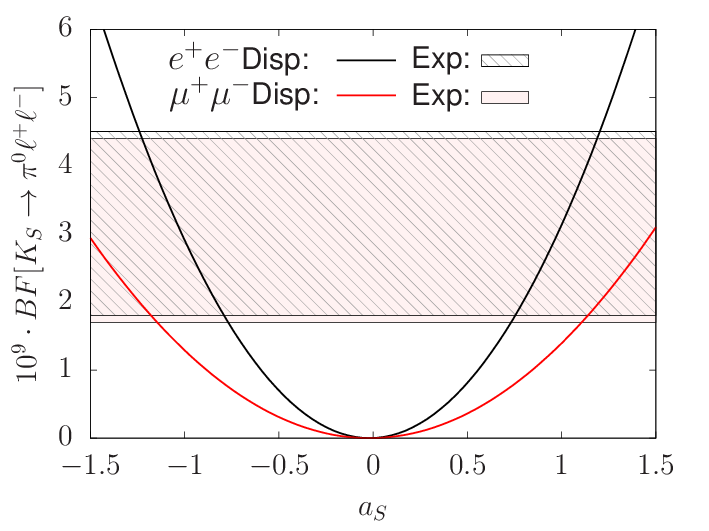}
\caption{\sl  Branching fractions for $K_S\to\piz e^+e^-$ ($m_{ee}>
  0.165$ GeV) and $K_S\to\piz \mu^+\mu^-$ as a function of $a_S$ 
  (with $a_+=-0.575$ fixed) in the  dispersive approach compared with the 
  experimental results~\cite{NA481:2003cfm,NA481:2004nbc}.}
\label{fig:a_Splot}
\end{figure}

\subsection{Results for $W_S(s)$}
We turn now to the amplitude $W_S$. Experimentally, two branching
fraction measurements are available, performed by the NA48 collaboration:  
a) of the mode  $K_S\to\pi^0e^+ e^-$ in the energy range 
$E_{ee}> 165$ MeV~\cite{NA481:2003cfm} and 
b) of the mode $K_S\to\pi^0\mu^+\mu^-$
(ref.~\cite{NA481:2004nbc}). We reproduce these two results below
\be\lbl{expW_S}
\ba{ll}
\left.BF(K_S\to\pi^0e^+e^-)\right\vert_{E_{ee}>165\,\hbox{MeV}}&= 
(3.0^{+1.5}_{-1.2}\pm0.2)\cdot10^{-9}\\
BF(K_S\to\pi^0\mu^+\mu^-) &= (2.9^{+1.5}_{-1.2}\pm0.2)\cdot10^{-9}
\ea
\en
and compare them with the branching fractions computed in the dispersive
approach varying the value of $a_S$ (with $a_+$ fixed) in fig.~\fig{a_Splot}.
One sees from the figure that there are two values of $a_S$ for which agreement 
with the experimental branching ratios can be achieved. Computing the $\chi^2$
one finds that the two minimums have approximately the same $\chi^2$ 
value, $\chi^2\simeq1.27$ and they correspond to the following results for $a_S$,
\be\lbl{a_Smini}
a_S=-1.15\pm0.20,\quad 
a_S=+1.10\pm0.20\ 
\en
which are in agreement with those estimated 
earlier~\cite{NA481:2003cfm,NA481:2004nbc,Buchalla:2003sj,Friot:2004yr}.
Computing the $\ell^+\ell^-$ branching fractions 
from the dispersive  amplitudes 
with either one of the two values of $a_S$  given in~\rf{a_Smini} 
gives approximately the same result
\be
\ba{ll}
\left.BF(K_S\to\pi^0e^+e^-)\right\vert_{E_{ee}>165\,\hbox{MeV}}& =(3.83\pm1.34)\cdot10^{-9}\\
BF(K_S\to\pi^0\mu^+\mu^-) &=(1.71\pm0.60)\cdot10^{-9}\ .
\ea\en
The curves in fig.~\fig{a_Splot} are nearly symmetric
with respect to $a_S=0$. Therefore, measuring the branching fractions with a
higher precision would provide a strong test of the approach but will not
allow to determine the sign of $a_S$.

\begin{figure}[hbt]
\centering  
\includegraphics[width=0.6\linewidth]{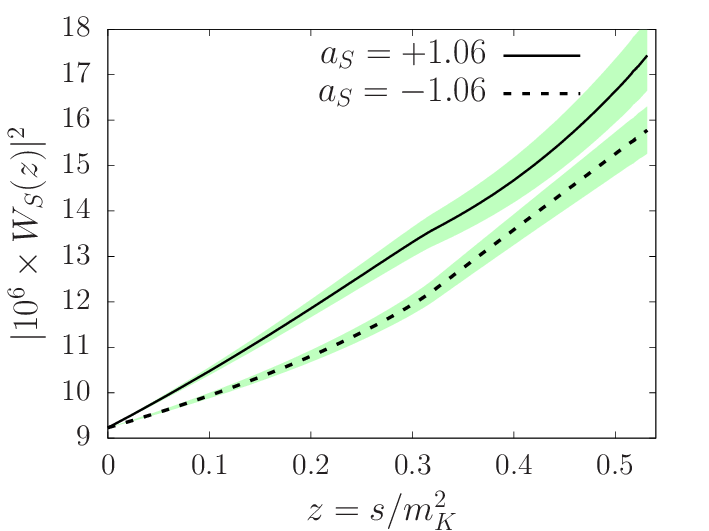}
\caption{\sl  Predicted dependence of $|W_S|^2$ as a function of $s$ for two different signs of $a_S$.}
\label{fig:W2_S}
\end{figure}

Some sensitivity to the sign of $a_S$ is exhibited in the energy dependence of
$|W_S(s)|^2$. The dispersive results for this quantity, taking $a_S\pm1.06$ 
and $a_+=-0.575$ are shown in fig.~\fig{W2_S}. Clearly, the
difference between the results corresponding to the two signs is not as 
marked as it was in the case of $W_+$ and $a_+$. 
However, if experimental results were available for a few 
energy bins, a one-parameter fit could be able to determine 
the sign of $a_S$. 

For small values
of the energy $s$, as before, the following 
linear approximation of 
$W_S$ holds: 
$W_S(s)\simeq G_{\rm F} [a_S\,\mkd + \bar{b}_S\,s ]$
and the slope parameter $\bar{b}_S$ is a linear function of $a_+$ and $a_S$
\be\lbl{W_Slinear}
\bar{b}_S =  0.259 +0.186\,a_+ + 0.500\,a_S .
\en
This formula shows that, assuming that $a_+\simeq -0.575$ 
and $|a_S|\gapprox1$, the sign of the
slope at the origin is the same as the sign of $a_S$ . 
The presence of the
term which is not proportional to $a_S$ in 
eq.~\rf{W_Slinear} allows, in principle, to distinguish
between the two sign possibilities based on the amplitude squared.

\subsection{Additional contributions: $\omega$ and $\phi$ resonances}
An approximate evaluation of the contributions from the 
$I=0$ resonances $\omega(782)$, $\phi(1020)$ to the form 
factors $W_{+,S}$ can be performed
assuming nonet symmetry and $\Delta{I}=1/2$ dominance. 
For this purpose, we start from the resonance-chiral
Lagrangian proposed in ref.~\cite{Ecker:1992de} which 
generalises to the
$\Delta{S}=1$ sector (and $\Delta{I}=1/2$) the formalism 
first introduced in
ref.~\cite{Ecker:1988te} for the strong, electromagnetic 
and semi-leptonic
sectors. In this formalism, vector and axial-vector resonances 
are described
with antisymmetric tensor fields. These can be ascribed a 
chiral order $p^2$
and the Lagrangian written in ref.~\cite{Ecker:1992de}
(we refer to this
article for details on the notation used below) contains a 
complete set
of independent terms which have a chiral order $p^4$. At this order, flavour
symmetry is left unbroken, such that the vector resonances 
are degenerate in mass
if we further assume nonet symmetry i.e. that double-trace terms are suppressed by the OZI
rule.  There are three relevant terms involving vector mesons which contribute
to the $W_{+,S}$ form factors\footnote{Unlike in the vector 
field formalism (e.g.~\cite{Dambrosio:1997ctq,Dubnickova:2006mk}) axial-vector resonances do not contribute to the $W_{+,S}$ 
form factors in the antisymmetric tensor field formalism at 
this chiral order.}, 
\be
\lbl{LagVDeltaS1}
{\cal L}_{\Delta{S}=1}(V)=
   g_V^1 \braque{\Delta\{ V_{\mu\nu},f_+^{\mu\nu}\}}
  +i g_V^5  \braque{\Delta \{V_{\mu\nu}, [u^\mu,u^\nu] \}}  
  +i g_V^6  \braque{\Delta   u_\mu V^{\mu\nu} u_\nu} \ ,
\en
where $\Delta= u\lambda_6 u^\dagger$ and 
we have also assumed that the OZI suppressed term
$\braque{V^{\mu\nu}}\braque{\Delta[u_\mu,u_\nu]}$ can be neglected. The
following terms from the ordinary sector are also involved,
\be
\ba{ll}
{\cal L}(V)= &-\frac{1}{2}\braque{\nabla^\lambda V_{\lambda\mu} 
\nabla_\nu V^{\nu\mu} -\frac{1}{2} M_V^2 V_{\mu\nu} V^{\mu\nu} }\\[0.3cm]
\ &+ \dfrac{F_V}{2\sqrt2} \braque{V_{\mu\nu} f^{\mu\nu}_+}
+\dfrac{iG_V}{\sqrt2} \braque{V_{\mu\nu} u^\mu u^\nu}
\ea\en
as well as terms from the $O(p^2)$ chiral Lagrangian
\be
{\cal L}^{(2)}= \frac{F_\pi^2}{4} \braque{u^\mu u_\mu +\chi^+},\quad
{\cal L}^{(2)}_{\Delta{S}=1}   =G_8F_\pi^4  \braque{\Delta u^\mu u_\mu} \ . 
\en
We can now compute the radiative amplitudes $K\to\pi\gamma^*$ from the
Lagrangian terms given above at tree level. The resulting amplitudes
are expressed  as a sum of poles. The $I=3/2$  combination has no
contributions from the $I=0$ resonances, it reads,
\be\lbl{W32reso}
\frac{W^{[3/2]}_{res}(s)}{16\pi^2\mkd}=
-\frac{\sqrt2 F_V}{\fpid}\frac{(g_V^6 +\sqrt2
G_8G_V\fpid)}{M_\rho^2-s}\ ,
\en
and the $I=1/2$ combination has the following expression
\be\lbl{W12reso}
\ba{ll}
\dfrac{W^{[1/2]}_{res}(s)}{16\pi^2\mkd}= &
\dfrac{4\sqrt2 G_V}{\fpid}\dfrac{g_V^1}{M^2_{K^*}-s}
+\dfrac{F_V}{\sqrt2\fpid}\Big[\dfrac{6g_V^5-g_V^6}{M^2_\rho-s}
-\dfrac{2g_V^5+g_V^6}{M^2_\omega-s}+\dfrac{4g_V^5}{M^2_\phi-s}\Big]\\[0.3cm]
\ & +\dfrac{G_8G_VF_V}{\mkd-\mpid}\Big[\dfrac{-4\mkd+\mpid}{M^2_\rho-s}
+\dfrac{\mpid}{M^2_\omega-s}+\dfrac{2\mpid}{M^2_\phi-s}\Big]\ .
\ea
\en
The $\rho(770)$ resonance corresponds to a pole on the second Riemann
sheet in the dispersive amplitudes. Equating the residues of this pole
in the $W_+$, $W_S$ amplitudes with those in the resonance model
allows to determine the values of the two couplings $g_V^5$,
$g_V^6$. Similarly the coupling $g_V^1$ can be estimated from the residue
of the $K^*(892)$ pole.
In practice, we have estimated these couplings more simply
by fitting the imaginary parts of the amplitudes in the resonance
model, after including physical values for the masses and the widths in the
denominators, to the imaginary parts of the dispersive amplitudes. The
resulting values of the couplings $g_V^i$ are 
collected\footnote{The following numerical
input values are used: $G_V=53.0$, $F_V=154.0$, $\fpi=92.2$ (all in MeV)
and $G_8=6.54\cdot10^{-6}$ $\hbox{GeV}^{-2}$.} in table~\ref{table:numvalgVi}.
\begin{table}
\centering
\bt{cccc}\hline\hline
\TT \BB $a_S$ & $10^8g_V^1/\fpi$ & $10^8g_V^5/\fpi$ & $10^8g_V^6/\fpi$\\ \hline
\TT $+1.06$ & $-3.38$ & $-0.06$ & $-5.91$ \\
$-1.06$ &  $0.15$ &  $2.24$ & $-1.42$ \\  \hline\hline   
\et
\caption{\small Approximate values of the three coupling constants
$g_V^1$,$g_V^5$,$g_V^6$ corresponding to two input values of $a_S$.  }
\label{table:numvalgVi}
\end{table}

The radiative amplitudes  being approximately linear in $s$ in the
physical decay region $[4m_l^2,(\mk-\mpi)^2]$, we can
characterise in a simple way the effect of the $I=0$ resonances $\omega,\phi$ 
from their contributions to the derivative at $s=0$, i.e. to the slope
parameter. Computing the derivative from eq.~\rf{W12reso} one
obtains, in the nonet symmetry limit,
\be\lbl{slopeomegphi}
G_F \bar{b}^{[1/2]}_{\omega,\phi}=\frac{16\pi^2\mkd F_V}{\sqrt2\fpid M^4_V}
\Big[2g_V^5-g_V^6 +\frac{3\sqrt2 G_8G_V\fpid\mpid}{\mkd-\mpid}\Big]\ .
\en
The corresponding slopes of $W_{+,S}$ are  simply related to
$\bar{b}^{[1/2]}_{\omega,\phi}$ by: 
$\bar{b}_+\vert_{\omega,\phi}=\bar{b}^{[1/2]}_{\omega,\phi}/3$,
$\bar{b}_S\vert_{\omega,\phi}=-\bar{b}^{[1/2]}_{\omega,\phi}/3$. 
One sees from the numbers given  
in table~\ref{table:numvalgVi} that the slopes
generated by $\omega,\phi$ are rather small and nearly independent of 
the sign of $a_S$:
\be
\bar{b}_+\vert_{\omega,\phi}\simeq+0.25\ (a_S>0),\quad
\bar{b}_+\vert_{\omega,\phi}\simeq+0.26\ (a_S<0)\ 
\en
using $M_V=M_\omega$.
This reflects the fact that $\im[W_+]$ (unlike $\re[W_+]$),
  from which the combination $2g_V^5-g_V^6$ is derived, has a very small
  dependence on $a_S$ at low energy and up to the $\rho$ mass.
  
\begin{figure}[htb]
\centering
\includegraphics[width=0.50\linewidth]{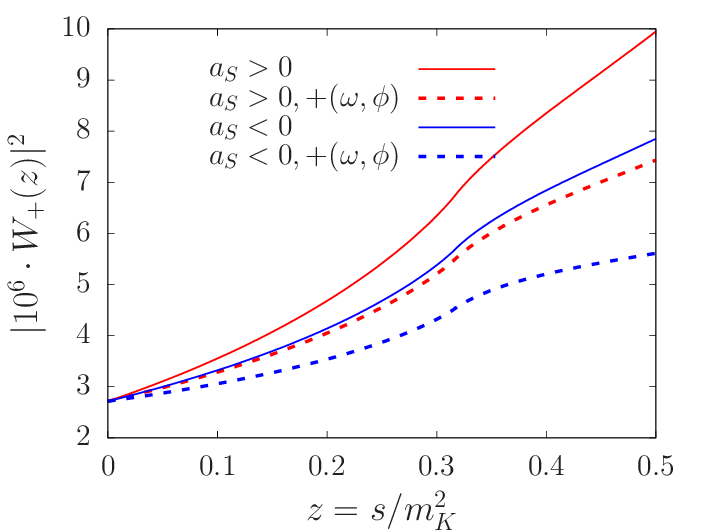}%
\includegraphics[width=0.50\linewidth]{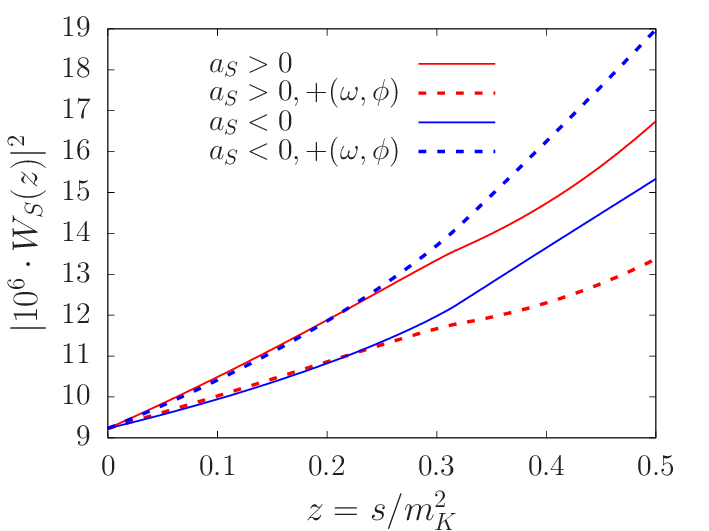}
\caption{\small Contributions of the $I=0$ resonances 
$\omega$, $\phi$
to $W_+$ (left) and $W_S$ (right). The values at $s=0$ are 
fixed with $a_+=-0.575$ and $a_S=\pm1.06$. The solid and dashed
red lines show the dispersive results when $a_S>0$ without (solid) 
and with (dashed) the $I=0$ resonances.
The solid and dashed blue lines show the results without and with the $I=0$ resonances when $a_S$ is negative. }
\label{fig:W2_gv5gv6}
\end{figure}
The results of
including the effects of the $I=0$ resonances, estimated in the way described
above, is illustrated in fig.~\fig{W2_gv5gv6}. In the case of $W_+$, 
they decrease the modulus squared of the
amplitudes in both cases of $a_S>0$ and $a_S<0$. The former case thus
remains favoured by comparison with the experimental data.
A similar effect of the $I=0$ resonances to $W_+$ 
was observed also in the factorisation approach~\cite{Okun:1975uq,Bergstrom:1990vj}.
In the case of $|W_S|^2$,  
the $I=0$ resonances
{modify significantly} the difference between 
the results corresponding to $a_S>0$ and $a_S<0$ 
(see fig.~\fig{W2_gv5gv6}, right).

\subsection{Additional contributions: higher mass resonances}
We can make a simple qualitative estimate of the role of 
higher mass resonances in the following way. In the 
dispersive representation~\rf{W12nu}
the contributions from the energy region above the cutoff 
$\Lambda$ are absorbed into a constant $C^{[1/2]}(\nu)$.  
As shown in~\cite{DAmbrosio:2018ytt}
the dependence on the scale $\nu$ of this constant
can be generated in a Regge-type model involving a sum over an
infinite set of resonances with equally spaced masses squared such that the
asymptotic behaviour in $s$ is of the 
form $\sim\log(s/\nu^2)$. The coefficient of this 
logarithmic function must coincide with the leading term in the
short-distance operator-product expansion of $T(j_\mu(x)\sum C_i O_i(y))$.
Using the model developed in~\cite{DAmbrosio:2018ytt}, and 
keeping the resonances which have a mass higher than 1 GeV 
in the sum amounts to replace 
$C^{[1/2]}(\nu)$ by a constant independent of $\nu$ plus a function
$C^{[1/2]}(s;\nu)$ of the following form
\be\lbl{sumreso}
C^{[1/2]}(s;\nu)=\frac{G_F}{\sqrt2}V_{ud}V_{us} \frac{8\mkd}{3}\,(N_c (C_1-C_4)+C_2-C_3)
\Big(\frac{5}{3}+\log\frac{\nu^2}{M^2}-\psi(3-\frac{s}{M^2})\,\Big)\ .
\en
The resonance poles generated by the $\psi$ function 
correspond to  masses
$M_n= M\sqrt{n+2}$ with $n\ge1$. Taking $M=0.87$ GeV as obtained in
~\cite{DAmbrosio:2024ncc} gives 
$M_n=1.51,1.74,1.95,\cdots$ GeV. Finally, the contribution
from the infinite sum over these higher mass states to 
the derivative at $s=0$ is
easily obtained from~\rf{sumreso},
\be
G_F \bar{b}^{[1/2]}_{M_n>\Lambda}= \frac{G_F}{\sqrt2}V_{ud}V_{us} \frac{8\mkd}{3M^2}
\,(N_c (C_1-C_4)+C_2-C_3)\, \psi'(3)\simeq -2.82\cdot10^{-3}
\en
using $C_1=-0.506$, $C_2=1.270$, $C_3=0.013$, $C_4=-0.034$ (values at 
a scale $\nu=1$ GeV, in the NDR scheme, taken from~\cite{Buchalla:1995vs})
and $\psi'(3)=\pi^2/6-5/4$. The
contribution to the slope parameter from these higher 
mass states, in this model, is seen to be essentially 
negligible being smaller
than the corresponding slope generated by the light 
resonances $\omega,\phi$ by two orders of magnitude. 

\section{Conclusions}
We have developed a description of the $K\to\pi\ell^+\ell^-$ 
amplitudes $W_+$
and $W_S$ in a dispersive theoretical approach which goes beyond previous
studies (e.g.~\cite{DAmbrosio:1998gur}) in taking explicitly 
into account, in
particular, the effects of the light vector resonances $\rho(770)$,
$K^*(892)$. This is achieved, on the one hand, by making use 
of new results on
the $K\to3\pi$ amplitudes~\cite{Bernard:2024ioq}, which have a domain of
validity extending beyond the physical decay region and, 
on the other hand, by
including the $K\pi$ states into the unitarity relations and 
the corresponding integral representations. The structure of these is
compatible with the requirement that the QCD scale dependence should cancel
upon adding the contribution associated with the operator $O_{7V}$.

Considering the two combinations $W^{[1/2]}\equiv2W_+-W_S$ 
and $W^{[3/2]}=W_++W_S$ we have investigated dispersive 
representations which 
involve the minimal number of subtraction constants compatible with their
asymptotic behaviours. The resulting coupled system of equations 
for $W_+$, $W_S$ are given in eqs.~\rf{W32} and~\rf{W12phys}
depends on only two parameters $a_+$ and $a_S$.

Other light vector resonances are present, in particular
$\omega(782),\phi(1019)$, but their influence on the $W_{+,S}$ amplitudes
cannot be accounted for in the same way as was done for the $\rho$ and the
$K^*$ resonances, because the corresponding discontinuities involve amplitudes,
$K\pi\to3\pi$, $K{\Kbar}\to{K}\pi$ which cannot be accessed
experimentally. We estimated their contributions in an approximate way
by appealing to nonet symmetry, $\Delta{I}=1/2$ dominance, and relying 
on a chiral-resonance Lagrangian, 
{finding them to be less important than the $I=1,1/2$
resonances but possibly significant. In a general way, the isoscalar
resonances can be accounted for model independently by introducing one
additional subtraction parameter  in the $W^{[1/2]}$ dispersive
representation. }

The main outcomes of our dispersive representations are illustrated in 
figs.~\fig{W2_plus} and~\fig{W2_S} showing the results for the amplitudes
squared $|W_+|^2$ and $|W_S|^2$ corresponding to different signs 
of $a_+$ and $a_S$. 
A positive sign for $a_+$ is seen to be clearly excluded in this
approach. Of more interest, perhaps, is the sign of $a_S$. Fig.~\fig{W2_S}
shows that the $s$ dependence of $|W_S|^2$ has a certain sensitivity to this
sign but it appears to be weakened upon including the effects of the $I=0$
resonances (see fig.~\fig{W2_gv5gv6}). However, $|W_+|^2$ is also sensitive 
to the sign of $a_S$, because of the coupling between $W_+$ and $W_S$,  as
one can see rather clearly in fig.~\fig{W2_plus}. 
Given some improved data on
$|W_S|^2$ one could hope to derive strong constraints both  on $|a_S|$
and its sign by performing a combined fit of the $|W_+|^2$ and $|W_S|^2$
data.  At the moment, we note that the curve which corresponds to 
the central values $a_+=-0.575$, $a_S=+1.06$ (taken simply 
from the literature) is in rather good agreement with the experimental 
data.

Another aspect of our analysis concerns the $\Delta{I}=1/2$ part of the
$K_S\to\pip\pim\piz$ amplitude, which has not been determined so far. In the
KT approach this part involves a single parameter $\tilde{\mu}_1$ 
and we have derived a linear relation between $\tilde{\mu}_1$ and 
the sum $a_+ +a_S$ (eq.~\rf{Wsum2s=0}) which allows to 
determine $\tilde{\mu}_1$. The
effect of this $\Delta{I}=1/2$ amplitude is chirally suppressed inside the
Dalitz plot but it is not excluded that it could be measured in the future.


\end{document}